\pdfoutput=1

\documentclass[acmlarge,natbib]{acmart}

\setcopyright{acmlicensed}
\acmJournal{TOIS}
\acmYear{2018} \acmVolume{36} \acmNumber{4} \acmArticle{38} \acmMonth{6} \acmPrice{$15.00$}\acmDOI{10.1145/3196826}

\usepackage{amsmath}
\usepackage{booktabs}
\usepackage{color}
\usepackage{enumitem}
\usepackage{flushend}
\usepackage{footnote}
\usepackage{hyperref}
\usepackage{grffile}
\usepackage{mathtools}
\DeclarePairedDelimiter{\ceil}{\lceil}{\rceil}
\usepackage{multirow}
\usepackage[np, autolanguage]{numprint}
\usepackage[nodisplayskipstretch]{setspace}
\usepackage{subfig}
\usepackage{tabularx}
\usepackage{tikz}
\usepackage{tikz-3dplot}
\usepackage{todonotes}
\usepackage[labelfont=bf,textfont=bf,skip=0pt]{caption}
\usepackage{etoolbox}
\usepackage{array}
\usepackage{xintfrac}
\usepackage[normalem]{ulem}

\newenvironment{inparaenum}[1][]{\begin{enumerate}}{\end{enumerate}}

\setcitestyle{square,comma,sort&compress}

\newcommand{\FullDocToVec}{doc2vec}
\newcommand{\FullWordToVec}{word2vec}
\newcommand{\DocToVec}{d2v}
\newcommand{\WordToVec}{w2v}
\newcommand{\CombineSelfInformation}{\texttt{si}}
\newcommand{\CombineAdd}{\texttt{add}}
\newcommand{\WordToVecSgSi}{\WordToVec{}-\CombineSelfInformation{}}

\newcommand{\LSI}{LSI}
\newcommand{\LDA}{LDA}
\newcommand{\QLM}{QLM}
\newcommand{\LSE}{LSE}
\newcommand{\ModelName}{NVSM}
\newcommand{\FullModelName}{Neural Vector Space Model}

\newcommand{\Dirichlet}{\texttt{d}}
\newcommand{\JelinekMercer}{\texttt{jm}}

\newcommand{\FullDirichlet}{Dirichlet}
\newcommand{\FullJelinekMercer}{Jelinek-Mercer}

\newcommand{\QLMDirichlet}{\QLM{} (\Dirichlet{})}
\newcommand{\QLMJelinekMercer}{\QLM{} (\JelinekMercer{})}

\newcommand{\TitlestatRelevant}{\texttt{titlestat\_rel}}
\newcommand{\MAPCut}{MAP@1000}
\newcommand{\NDCGCut}{NDCG@100}
\newcommand{\Precision}{P@10}

\newcommand{\BenchmarkAP}{Associated Press 88-89}
\newcommand{\BenchmarkWSJ}{Wall Street Journal}
\newcommand{\BenchmarkLATimes}{LA Times}
\newcommand{\BenchmarkFT}{Financial Times}
\newcommand{\BenchmarkRobust}{Robust04}
\newcommand{\BenchmarkNYT}{New York Times}

\newcommand{\ShortBenchmarkAP}{AP88-89}
\newcommand{\ShortBenchmarkWSJ}{WSJ}
\newcommand{\ShortBenchmarkLATimes}{LA}
\newcommand{\ShortBenchmarkFT}{FT}
\newcommand{\ShortBenchmarkRobust}{\BenchmarkRobust{}}
\newcommand{\ShortBenchmarkNYT}{NY}

\newcommand{\Length}[1]{|#1|}

\newcommand{\Vocabulary}{V}

\newcommand{\Queries}{Q}
\newcommand{\Query}{\MakeLowercase{\Queries{}}}
\newcommand{\QueryTerm}[1]{t_{#1}}

\newcommand{\Documents}{D}
\newcommand{\Document}{\MakeLowercase{\Documents{}}}
\newcommand{\Word}{w}

\newcommand{\LatentDimension}[1]{{k_{#1}}}
\newcommand{\EmbeddingMatrix}[1]{{R_{#1}}}

\newcommand{\WordMatrix}{\EmbeddingMatrix{\Vocabulary{}}}

\newcommand{\LatentWordDim}{\LatentDimension{\Word{}}}

\newcommand{\DocumentMatrix}{\EmbeddingMatrix{\Documents{}}}

\newcommand{\LatentDocumentDim}{\LatentDimension{\Document{}}}

\newcommand{\TransformFn}{f}
\newcommand{\TransformMatrix}{W}

\newcommand{\NGramSize}{n}

\newcommand{\CompositionalFn}{g}

\newcommand{\HashFn}{h}

\newcommand{\ScoreFn}{\text{score}}

\newcommand{\Apply}[2]{{#1\left(#2\right)}}
\newcommand{\ApplySquare}[2]{{#1\left[#2\right]}}
\newcommand{\Transpose}[1]{{{#1}^\intercal}}

\newcommand{\Prob}[1]{\Apply{P}{#1}}
\newcommand{\CondProb}[2]{\Prob{#1 \mid #2}}

\newcommand{\ApproxProb}[1]{\tilde{P}\left(#1\right)}
\newcommand{\CondApproxProb}[2]{\ApproxProb{#1 \mid #2}}

\newcommand{\DotProduct}[2]{#1 \cdot #2}

\newcommand{\SampleExpectation}{\hat{\mathbb{E}}}
\newcommand{\SampleVariance}{\hat{\mathbb{V}}}

\newcommand{\ErrorBias}{\beta}
\newcommand{\NGramProjection}{\Apply{\Projection{}}{\BatchInstancePhrase{i}}}
\newcommand{\HardTanH}{\text{hard-tanh}}

\newcommand{\BenchmarkFigureWidth}{0.3225\textwidth}

\def \paperImplementationUrl {\url{https://github.com/cvangysel/cuNVSM}}

\def \ResearchQuestionOne {How does \ModelName{} compare to other latent vector space models, such as \FullDocToVec{} \citep{Le2014}, \FullWordToVec{} \citep{Mikolov2013word2vec,Vulic2015monolingual}, \LSI{} \citep{Deerwester1990}, \LDA{} \citep{Blei2003} and \LSE{} \citep{VanGysel2016products}, on the document retrieval task?}
\def \ResearchQuestionTwo {For what proportion of queries does \ModelName{} perform better than the other rankers?}
\def \ResearchQuestionThree {What gains does \ModelName{} bring when combined with a lexical \QLM{} and a competing state-of-the-art vector space model?}
\def \ResearchQuestionFour {Do \ModelName{}s exhibit regularities that we can link back to well-understood document collection statistics used in traditional retrieval models and how do the regularities in \ModelName{} differ from those in \LSE{}?}

\def \ContributionOne {A novel neural retrieval model, \ModelName{}, that is learned using gradient descent on a document collection.}
\def \ContributionTwo {Comparisons of lexical (\QLM{}) and semantic (\FullDocToVec{}, \FullWordToVec{}, \LSI{}, \LDA{}, \LSE{} and \ModelName{}) models on document retrieval test collections.}
\def \ContributionThree {An analysis of the internals of \ModelName{} to give insights in the workings of the model and the retrieval task.}
\def \ContributionFour {A highly-optimized open source C++/CUDA implementation of \ModelName{} that results in fast training and efficient memory usage.}
\def \ContributionFive {Advice on how to configure the hyperparameters of \ModelName{}.}

\newcommand{\EuclideanNorm}[1]{\left\lVert#1\right\rVert}

\newcommand{\abs}[1]{|{#1}|}

\makeatletter
\newcolumntype{"}{@{\hskip\tabcolsep\vrule width 1pt\hskip\tabcolsep}}
\makeatother

\newcommand{\splitatcommas}[1]{%
  \begingroup
  \begingroup\lccode`~=`, \lowercase{\endgroup
    \edef~{\mathchar\the\mathcode`, \penalty0 \noexpand\hspace{0pt plus 1em}}%
  }\mathcode`,="8000 #1%
  \endgroup
}

\allowdisplaybreaks

\begin{document}

\title{Neural Vector Spaces for Unsupervised Information Retrieval}

\author{Christophe Van Gysel}
\orcid{0000-0003-3433-7317}
\email{cvangysel@uva.nl}
\affiliation{%
\institution{University of Amsterdam}
\city{Amsterdam}
\country{The Netherlands}
}

\author{Maarten de Rijke}
\orcid{0000-0002-1086-0202}
\email{derijke@uva.nl}
\affiliation{%
\institution{University of Amsterdam}
\city{Amsterdam}
\country{The Netherlands}
}

\author{Evangelos Kanoulas}
\orcid{}
\email{e.kanoulas@uva.nl}
\affiliation{%
\institution{University of Amsterdam}
\city{Amsterdam}
\country{The Netherlands}
}

\begin{abstract}
We propose the \FullModelName{} (\ModelName{}), a method that learns representations of documents in an unsupervised manner for news article retrieval. In the \ModelName{} paradigm, we learn low-dimensional representations of words and documents from scratch using gradient descent and rank documents according to their similarity with query representations that are composed from word representations. We show that \ModelName{} performs better at document ranking than existing latent semantic vector space methods. The addition of \ModelName{} to a mixture of lexical language models and a state-of-the-art baseline vector space model yields a statistically significant increase in retrieval effectiveness. Consequently, \ModelName{} adds a complementary relevance signal. Next to semantic matching, we find that \ModelName{} performs well in cases where lexical matching is needed.
\ModelName{} learns a notion of term specificity directly from the document collection without feature engineering. We also show that \ModelName{} learns regularities related to Luhn significance. Finally, we give advice on how to deploy \ModelName{} in situations where model selection (e.g., cross-validation) is infeasible. We find that an unsupervised ensemble of multiple models trained with different hyperparameter values performs better than a single cross-validated model. Therefore, \ModelName{} can safely be used for ranking documents without supervised relevance judgments.
\end{abstract}




\thanks{This research was supported by
Ahold Delhaize,
Amsterdam Data Science,
the Bloomberg Research Grant program,
the Criteo Faculty Research Award program,
Elsevier,
the European Community's Seventh Framework Programme (FP7/2007-2013) under
grant agreement nr 312827 (VOX-Pol),
the Google Faculty Research Award scheme,
the Microsoft Research Ph.D.\ program,
the Netherlands Institute for Sound and Vision,
the Netherlands Organisation for Scientific Research (NWO)
under pro\-ject nrs
612.\-001.\-116, % ImFIRE
CI-14-25, % MediaNow
652.\-002.\-001, % Re-Search
612.\-001.\-551, % CLEAR
652.\-001.\-003, % NEWEL
and
Yandex.
All content represents the opinion of the authors, which is not necessarily shared or endorsed by their respective employers and/or sponsors.
}

\maketitle

% !TEX root = ./tois2018-adhoc-retrieval.tex

\section{Introduction}

The vocabulary mismatch between query and document poses a critical challenge in search \citep{Li2014}. The vocabulary gap occurs when documents and queries, represented as a bag-of-words, use different terms to describe the same concepts. While improved semantic matching methods are urgently needed, in order for these methods to be effective they need to be applicable at early stages of the retrieval pipeline. Otherwise, candidate documents most affected by the mismatch (i.e., relevant documents that do not contain any query terms) will simply remain undiscovered. \citet{Boytsov2016knn} show that (approximate) nearest neighbor algorithms \citep{Garcia2008fastknn,muja2014scalableknn} can be more efficient than classical term-based retrieval. This strongly motivates the design of semantic matching methods that represent queries and documents in finite-dimensional vector spaces.

Latent semantic models, such as \LSI{} \citep{Deerwester1990}, fit in the finite-dimen\-sional vector space paradigm needed for nearest neighbor retrieval. However, \LSI{} is known to retrieve non-relevant documents due to a lack of specificity \citep{Dumais95lsi}. The recent move towards learning word representations as part of neural language models \citep{Bengio2003} has shown impressive improvements in natural language processing (NLP) \citep{Collobert2011scratch,Mikolov2013word2vec,Graves2014speech}. Therefore, it is reasonable to explore these representation learning methods for information retrieval (IR) as well. Unfortunately, in the case of full text document retrieval, only few positive results have been obtained so far \citep{Craswell2016neuir}. We identify two causes for this shortfall.

First of all, IR tasks (e.g., document ranking) are fundamentally different from NLP tasks \citep{Craswell2016neuir}. NLP deals with natural language regularities (e.g., discovering long range dependencies), whereas IR involves satisfying a user's information need (e.g., matching a query to a document). Therefore, specialized solutions and architectures for IR are needed.

Secondly, in the bag-of-words paradigm, query/document matching is performed by counting term occurrences within queries and documents. Afterwards, the frequencies are adjusted using weighting schemes that favor term specificity \citep{Robertson1994bm25,Robertson2004idf}, used as parameter values in probabilistic frameworks \citep{Zhai2004smoothing} and/or reduced in dimensionality \citep{Blei2003,Deerwester1990}. However, \citet{Baroni2014} noted that for NLP tasks, prediction-based models, learned from scratch using gradient descent, outperform count-based models. Similarly, the advances made by deep learning in computer vision \citep{Krizhevsky2012net} were not due to counting visual words \citep{Vidal2003sift} or by constructing complicated pipelines, but instead by optimizing a cost function using gradient descent and learning the model from scratch. However, for IR settings such as unsupervised news article retrieval, where one ranks documents in the absence of explicit or implicit relevance labels, prediction-based models have not received much attention. In order for deep learning to become feasible for unsupervised retrieval, we first need to construct an appropriate optimization objective.

In this paper we introduce an optimization objective for learning latent representations of words and documents from scratch, in an unsupervised manner without relevance signals. Specifically, we introduce the \FullModelName{} (\ModelName{}) for document retrieval. The optimization objective of \ModelName{} mandates that word sequences extracted from a document should be predictive of that document. Learning a model of content in this way incorporates the following IR regularities:
\begin{itemize}
	\item \textbf{semantic matching}: words occurring in each other's vicinity are learned to have similar representations,
	\item the \textbf{clustering hypothesis}: documents that contain similar language will have nearby representations in latent space, and
	\item \textbf{term specificity}: words associated with many documents will be neglected, due to low predictive power.
\end{itemize}

One limitation of latent document vector spaces, including the \ModelName{} we introduce here, is that their asymptotic complexity is bounded by the number of documents (i.e., one vector for every document) \citep{Ai2016doc2veclm,VanGysel2016products,Chen2017corruption}. Consequently, the latent methods we consider in this paper are only feasible to be constructed on document collections of medium scale. Therefore, we choose to evaluate our methods on article retrieval benchmarks (${\sim}200\text{k}$ to $2\text{m}$ documents each) from TREC \citep{Zhai2004smoothing,Harman1992tipster}.

\clearpage
We seek to answer the following research questions:
\begin{inparaenum}[(1)]
	\item \ResearchQuestionOne{}
	\item \ResearchQuestionTwo{}
	\item \ResearchQuestionThree{}
	\item \ResearchQuestionFour{}
\end{inparaenum}

We contribute:
\begin{inparaenum}[(1)]
	\item \ContributionOne{}
	\item \ContributionTwo{}
	\item \ContributionThree{}
	\item \ContributionFour{}\footnote{\paperImplementationUrl{}\label{fn:ImplementationURL}}
	\item \ContributionFive{}
\end{inparaenum}
% !TEX root = ./tois2018-adhoc-retrieval.tex

\section{Related Work}

The majority of retrieval models represent documents and queries as elements of a finite-dimensional vector space and rank documents according to some similarity measure defined in this space. A distinction is made between bag-of-words models \citep{Sparck1972specificity,Robertson1994bm25,Zhai2004smoothing}, for which the components of the vector space correspond to terms, and latent models \citep{Deerwester1990,Hofmann1999,Blei2003,Ai2016doc2veclm,VanGysel2016products}, where the components are unobserved and the space is of low dimensionality compared to the number of unique terms.
We first discuss prior work on representation learning for document retrieval, with an emphasis on neural methods. Next, we give an overview of neural language modeling for natural language processing and automatic speech recognition.

\subsection{Unsupervised representations for IR}
\label{sec:related:unsupervised}

The recent revival of neural networks due to advances in computer vision \citep{Krizhevsky2012net}, natural language processing (NLP) \citep{Collobert2011scratch,Mikolov2013word2vec} and automatic speech processing (ASR) \citep{Graves2014speech} has led to an increasing interest in these technologies from the information retrieval community. In particular, word embeddings \citep{Bengio2003}, low-dimensional representations of words learned as part of a neural language model, have attracted much attention due to their ability to encode semantic relations between words. First we discuss non-neural latent semantic models, followed by work that integrates pre-trained neural embeddings into existing retrieval models, and work on neural retrieval models that are learned from scratch.

\subsubsection{Latent semantic models}
Latent Semantic Indexing (LSI) \citep{Deerwester1990} and probabilistic LSI (pLSI) \citep{Hofmann1999} were introduced in order to mitigate the mismatch between documents and queries \citep{Li2014}. \citet{Blei2003} proposed Latent Dirichlet Allocation (LDA), a topic model that generalizes to unseen documents.

\subsubsection{Combining pre-trained word or document representations}
The following methods take pre-existing representations and use them to either
\begin{inparaenum}[(1)]
	\item obtain a document representation from individual word representations that are subsequently used for (re-)ranking, or
	\item combine representation similarities in some way to (re-)rank documents.
\end{inparaenum}
While many of these methods were applied to word2vec \citep{Mikolov2013word2vec} representations at the time of publishing, these methods are in fact generic in the sense that they can be applied to any model that provides representations of the required unit (i.e., words or documents). Consequently, most of the methods below can be applied to \ModelName{} and the baseline vector space models we consider in this paper. However, the key focus of this paper is the alleviation of the vocabulary gap in information retrieval, and consequently for practical reasons, nearest neighbor search is the most viable way to compare latent vector space models \citep{Boytsov2016knn}.

\citet{Vulic2015monolingual} are the first to aggregate word embeddings learned with a context-predicting distributional semantic model (DSM); query and document are represented as a sum of word embeddings learned from a pseudo-bilingual document collection with a Skip-gram model.
\citet{Kenter2015shorttext} extract features from embeddings for the task of determining short text similarity. \citet{Zuccon2015nntm} use embeddings to estimate probabilities in a translation model that is combined with traditional retrieval models (similar to \citep{Ganguly2015generalizedlm,Tu2016semantic}). \citet{Zamani2016queryexpansion,Zamani2016embeddinglm} investigate the use of pre-trained word embeddings for query expansion and as a relevance model to improve retrieval. \citet{Guo2016wordtransport} introduce the Bag-of-Word-Embeddings (BoWE) representation where every document is represented as a matrix of the embeddings occurring in the document; their non-linear word transportation model compares all combinations of query/document term representations at retrieval time. They incorporate lexical matching into their model by exactly comparing embedding vector components for specific terms (i.e., specific terms occurring in both document and query are matched based the equality of their vector components, contrary to their lexical identity).

\subsubsection{Learning from scratch}

The methods discussed above incorporate features from neural language models. The recent deep learning revival, however, was due to the end-to-end optimization of objectives and representation learning \citep{LeCun1998gradient,Krizhevsky2012net,Sutskever2014seq2seq} in contrast to feature engineering or the stacking of independently-estimated models. The following neural methods learn representations of words and documents from scratch. \citet{Salakhutdinov2009} introduce semantic hashing for the document similarity task. \citet{Le2014} propose \FullDocToVec{}, a method that learns representations of words and documents. \citet{Ai2016doc2veclm} evaluate the effectiveness of \FullDocToVec{} representations for ad-hoc retrieval, but obtain disappointing results that are further analyzed in \citep{Ai2016doc2vecanalysis}. An important difference between the work of \citeauthor{Ai2016doc2veclm} and this paper is that we operate within the vector space framework. This allows us to retrieve documents using an (approximate) nearest neighbor search and thus mitigate the vocabulary gap at early stages of the retrieval pipeline. In contrast, the approach by \citet{Ai2016doc2veclm} can only be feasibly applied in a re-ranking setting. \citeauthor{VanGysel2016products} develop end-to-end neural language models for entity ranking tasks \citep{VanGysel2017sert}; they show how their neural models outperform traditional language models on the expert finding task~\citep{VanGysel2016experts} and improve the scalability of their approach and extend it to product search \citep{VanGysel2016products}.

\subsection{Neural IR and language modeling}

Beyond representation learning, there are more applications of neural models in IR \citep{Craswell2016neuir,Onal2017neural,Kenter2017nn4ir,VanGysel2017replywith}. In machine-learned ranking \citep{Liu2011}, we have RankNet \citep{Burges2005ranknet}. In the class of supervised learning-to-match approaches, where clicks are available, there are DSSM~\citep{Huang2013,Shen2014} and DSN~\citep{Deng2013}. \citet{Guo2016relevance} learn a relevance model by extracting features from BoWE representations in addition to corpus statistics such as inverse document frequency. Recently, \citet{Mitra2017distributed} have introduced a supervised document ranking model that matches using both local and distributed representations. Next to retrieval models there has been work on modeling user interactions with neural methods. \citet{Borisov2016click} introduce a neural click model that represents user interactions as a vector representation; in \citep{Borisov2016contextaware}, they extend their work by taking into account click dwell time.

Neural Network Language Models (NNLM) \citep{Bengio2003} have shown promising results in NLP \citep{Sutskever2014seq2seq,Jozefowicz2015rnn,Tran2016rmn} and ASR \citep{Graves2014speech,Sak2015acousticlstm} compared to Markovian models. \citet{Collobert2011scratch} apply NNLMs to arbitrary NLP tasks by learning one set of word representations in a multi-task setting. \citet{Turian2010representations} compare word representations learned by neural networks, distributional semantics and cluster-based methods as features for NLP tasks \citep{Baroni2014}. Following this observation, models have been designed with the sole purpose of learning embeddings \citep{Mikolov2013word2vec,Pennington2014,Kenter2016siamese}. \citet{Levy2014neural} show the relation between neural word embeddings and distributional word representations and how to improve the latter using lessons learned from the former \citep{Levy2015improving}.

\smallskip\noindent%
The contribution of this paper over and above the related work discussed above is the following.
First, we substantially extend the \LSE{} model of \citet{VanGysel2016products} and evaluate it on document search collections from TREC \citep{TRECAdhoc}. Our improvements over the \LSE{} model are due to
\begin{inparaenum}[(1)]
	\item increased regularization, and
	\item accelerated training by reducing the internal covariate shift, and
	\item the incorporation of term specificity within the learned word representations.
\end{inparaenum}

Second, we avoid \emph{irrational exuberance} known to plague the deep learning field \citep{Craswell2016neuir} by steering away from non-essential depth. That is, while we extend algorithms and techniques from deep learning, the model we present in this paper is shallow.

Third, and contrary to previous work where information from pre-trained word embeddings is used to enrich the query/document representation of count-based models, our model, \ModelName{}, is learned directly from the document collection without explicit feature engineering. We show that \ModelName{} learns regularities known to be important for document retrieval from scratch. Our semantic vector space outperforms lexical retrieval models on some benchmarks. However, given that lexical and semantic models perform different types of matching, our approach is most useful as a supplementary signal to these lexical models.

% !TEX root = ./tois2018-adhoc-retrieval.tex

\section{Learning Semantic Spaces}

In this section we provide the details of \ModelName{}. First, we give a birds-eye overview of the model and its parameters and explain how to rank documents for a given query. Secondly, we outline our training procedure and optimization objective. We explain the aspects of the objective that make it work in a retrieval setting. Finally, we go into the technical challenges that arise when implementing the model and how we solved them in our open-source release.

\subsection{The \FullModelName{}}

Our work revolves around unsupervised ad-hoc document retrieval where a user wishes to retrieve documents (e.g., articles) as to satisfy an information need encoded in query $\Query{}$. 

Below, a query $\Query{}$ consists of terms (i.e., words) $\QueryTerm{1}, \ldots, \QueryTerm{\Length{\Query{}}}$ originating from a vocabulary $\Vocabulary{}$, where $\Length{\cdot}$ denotes the length operator; $\Documents{}$ denotes the set of documents $\{\Document{}_1, \dots, \Document{}_{\Length{\Documents{}}}\}$. Every document $\Document{}_i \in \Documents{}$ consists of a sequence of words $\Word{}_{i,1}, \ldots, \Word{}_{i, \Length{\Document{}_i}}$ with $\Word{}_{i, j} \in \Vocabulary{}$, for all $1 \leq j \leq \Length{\Document{}_i}$.

We extend the work of \citet{VanGysel2016products} who learn low-dimen\-sional representations of words, documents and the transformation between them from scratch. That is, instead of counting term frequencies for use in probabilistic frameworks \citep{Zhai2004smoothing} or applying dimensionality reduction to term frequency vectors \citep{Deerwester1990,Wei2006lda}, we learn representations directly by gradient descent from sampled n-gram/document pairs extracted from the corpus.\footnote{Note here that the intention is not to use n-grams as a substitute for queries, but rather that n-grams (i.e., multiple words) contain more meaning (i.e., semantics) than a single word. The training procedure can thus be characterized as projecting the semantic representation of an n-gram close to the document from which the n-gram was extracted.} These representations are embedded parameters in matrices $\DocumentMatrix{} \in \mathbb{R}^{\Length{\Documents{}} \times \LatentDocumentDim{}}$ and $\WordMatrix{} \in \mathbb{R}^{\Length{\Vocabulary{}} \times \LatentWordDim{}}$ for documents $\Documents{}$ and vocabulary words $\Vocabulary{}$, respectively, such that $\vec{R}_\Vocabulary^{(i)}$ ($\vec{R}_\Documents^{(j)}$, respectively) denotes the $\LatentWordDim{}$-dimensional ($\LatentDocumentDim{}$-dimensional, respectively) vector representation of word $\Vocabulary{}_i$ (document $\Document{}_i$, respectively).

As the word representations $\vec{R}_\Vocabulary^{(i)}$ and document representations $\vec{R}_\Documents^{(j)}$ are of different dimensionality, we require a transformation $\TransformFn{}: \mathbb{R}^\LatentWordDim{} \to \mathbb{R}^\LatentDocumentDim{}$ from the word feature space to the document feature space. In this paper we take the transformation to be linear:
\begin{equation}
\label{eq:transform}
\TransformFn{}\left(\vec{x}\right) = \TransformMatrix{} \vec{x},
\end{equation}
where $\vec{x}$ is a $\LatentWordDim{}$-dimensional vector and $\TransformMatrix{}$ is a $\LatentDocumentDim{} \times \LatentWordDim{}$ parameter matrix that is learned using gradient descent (in addition to representation matrices $\WordMatrix{}$ and $\DocumentMatrix{}$).

\newcommand{\VerboseNGram}[1][\NGramSize{}]{\Word{}_1, \ldots, \Word{}_#1}

We compose a representation of a sequence of $\NGramSize{}$ words (i.e., an n-gram) $\VerboseNGram{}$ by averaging its constituent word representations:
\begin{equation}
\label{eq:averaging}
\CompositionalFn{}\left( \VerboseNGram{} \right) = \frac{1}{\NGramSize{}} \sum^{\NGramSize{}}_{i=1} 
\vec{R}_\Vocabulary^{(\Word{}_i)}.
\end{equation}
A query $\Query{}$ is projected to the document feature space by the composition of $\TransformFn{}$ and $\CompositionalFn{}$: $\Apply{\left(\TransformFn{} \circ \CompositionalFn{}\right)}{ \Query{}} = \Apply{\HashFn{}}{\Query{}}$. 

The matching score between a document $\Document{}$ and query $\Query{}$ is then given by the cosine similarity between their representations in document feature space:
\begin{equation}
\label{eq:score}
\Apply{\ScoreFn{}}{\Query{}, \Document{}} = \frac{\DotProduct{\Transpose{\Apply{\HashFn{}}{\Query{}}}}{{\vec{R}_\Documents^{(\Document)}}}}{ \EuclideanNorm{\Apply{\HashFn{}}{\Query{}}} \EuclideanNorm{\vec{R}_\Documents^{(\Document)}}} .
\end{equation}
We then proceed by ranking documents $\Document{} \in \Documents{}$ in decreasing order of $\Apply{\ScoreFn{}}{\Query{}, \Document{}}$ (see Eq.~\ref{eq:score}) for a given query $\Query{}$. Note that ranking according to cosine similarity is equivalent to ranking according to inverse Euclidean distance if the vectors are normalized. Therefore, ranking according to Eq.~\ref{eq:score} can be formulated as an (approximate) nearest neighbor search problem in a metric space.

The model proposed here, \ModelName{}, is an extension of the \LSE{} model \citep{VanGysel2016products}. \ModelName{}/\LSE{} are different from existing unsupervised neural retrieval models learned from scratch due to their ability to scale to collections larger than expert finding collections \citep{VanGysel2016experts} (i.e., ${\sim}10\text{k}$ entities/documents) and the assumption that words and documents are embedded in different spaces \citep{Le2014}. Compared to word embeddings \citep{Mikolov2013word2vec,Pennington2014}, \ModelName{}/\LSE{} learn document-specific representations instead of collection-wide representations of language; see \citep[p. 4]{VanGysel2016products} for a more in-depth discussion. The algorithmic contribution of \ModelName{} over \LSE{} comes from improvements in the objective that we introduce in the next section.

\subsection{The objective and its optimization}
\label{sec:methodology:objective}

\newcommand{\Batch}{B}
\newcommand{\BatchSize}{m}
\newcommand{\BatchInstance}[1]{\Batch{}_{#1}}
\newcommand{\BatchInstancePhrase}[1]{\BatchInstance{#1}^{(p)}}
\newcommand{\BatchInstanceDocument}[1]{\BatchInstance{#1}^{(d)}}

\newcommand{\LossFn}{L}
\newcommand{\Parameters}{\Theta}

\newcommand{\Ideal}[1]{{#1}^*}

\newcommand{\NormFn}{\text{norm}}

We learn representations of words and documents using mini-batches of $\BatchSize{}$ n-gram/document pairs such that an n-gram representation---composed out of word representations---is projected nearby the document that contains the n-gram, similar to \LSE{} \citep{VanGysel2016products}. The word and n-gram representations, and the transformation between them are all learned simultaneously. This is in contrast to representing documents by a weighted sum of pre-trained representations of the words that it contains \citep{Vulic2015monolingual}. A mini-batch $\Batch{}$ is constructed as follows:
\begin{inparaenum}[(1)]
	\item Stochastically sample document $\Document{} \in \Documents{}$ according to $\Prob{\Documents{}}$. In this paper, we assume $\Prob{\Documents{}}$ to be uniform, similar to \citep{Zhai2004smoothing}. Note that $\Prob{\Documents{}}$ can be used to incorporate importance-based information (e.g., document length).
	\item Sample a phrase of $\NGramSize{}$ contiguous words $\Word{}_1, \ldots{}, \Word{}_\NGramSize{}$ from document $\Document{}$.
	\item Add the phrase-document pair $\left(\Word{}_1, \ldots{}, \Word{}_\NGramSize{} ; \Document{} \right)$ to the batch.
	\item Repeat until the batch is full.
\end{inparaenum}

Given a batch $\Batch{}$, we proceed by constructing a differentiable, non-convex loss function $\LossFn{}$ of the parameters $\Parameters{}$ (e.g., $\WordMatrix{}, \DocumentMatrix{}$) and the parameter estimation problem is cast as an optimization problem that we approximate using stochastic gradient descent.

Denote $\BatchInstance{i}$ as the $i$-th n-gram/document pair of batch $\Batch{}$ and $\BatchInstancePhrase{i}$ ($\BatchInstanceDocument{i}$, respectively) as the n-gram (document, respectively) of pair $\BatchInstance{i}$. Further, we introduce an auxiliary function that L2-normalizes a vector of arbitrary dimensionality:
\begin{equation*}
\Apply{\NormFn{}}{\vec{x}} = \frac{\vec{x}}{\EuclideanNorm{\vec{x}}}.
\end{equation*}
\newcommand{\Projection}{T}%
\newcommand{\RawProjection}{\tilde{\Projection{}}}%
For an n-gram/document pair $\BatchInstance{i}$, the non-standardized projection of the n-gram into the $\LatentDocumentDim{}$-dimensional document feature space is as follows:
\begin{equation}
\label{eq:raw_projection}
\Apply{\RawProjection{}}{\BatchInstancePhrase{i}} = \Apply{\left(\TransformFn{} \circ \NormFn{} \circ \CompositionalFn{}\right)}{\BatchInstancePhrase{i}}.
\end{equation}
A few quick comments are in order. The function $\CompositionalFn{}$ (see Eq.~\ref{eq:averaging}) constructs an n-gram representation by averaging the word representations (embedded in the $\WordMatrix{}$ parameter matrix). This allows the model to learn semantic relations between words for the purpose of \textbf{semantic matching}. That is, the model does not learn from individual words, but instead it learns from an unordered sequence (order is not preserved as we sum in Eq.~\ref{eq:averaging}) of words that constitute meaning. As documents contain multiple n-grams and n-grams are made up of multiple words, semantic similarity between words and documents is learned. In addition, the composition function $\CompositionalFn{}$ in combination with L2-normalization $\Apply{\NormFn{}}{\cdot}$ causes words to compete in order to contribute to the resulting n-gram representation. Given that we will optimize the $\NGramSize{}$-gram representation to be close to the corresponding document (as we will explain below), words that are discriminative for the document in question will learn to contribute more to the n-gram representation (due to their discriminative power), and consequently, the L2-norm of the representations of discriminative words will be larger than the L2-norm of non-discriminative words. This incorporates a notion of \textbf{term specificity} into our model.

\newcommand{\BatchMean}{\ApplySquare{\SampleExpectation{}}{\Apply{\RawProjection{}}{\BatchInstancePhrase{i}}}}
\newcommand{\BatchVariance}{\ApplySquare{\SampleVariance{}}{\Apply{\RawProjection{}}{\BatchInstancePhrase{i}}}}

We then estimate the per-feature sample mean and variance
\[
\BatchMean{} 
\text{ and }
\BatchVariance{}
\]
over batch $\Batch{}$.
The standardized projection of n-gram/document pair $\BatchInstance{i}$ can then be obtained as follows:
\begin{equation}
\label{eq:projection}
\NGramProjection{} = \Apply{\HardTanH}{\frac{\Apply{\RawProjection{}}{\BatchInstancePhrase{i}} - \BatchMean{}}{\sqrt{\BatchVariance{}}} + \ErrorBias{}},
\end{equation}
where $\ErrorBias{}$ is a $\LatentDocumentDim{}$-dimensional bias vector parameter that captures doc\-ument-independent regularities corresponding to word frequency. While vector $\ErrorBias{}$ is learned during training, it is ignored during prediction (i.e., a nuisance parameter) similar to the position bias in click models \citep{Chapelle2009dbn} and score bias in learning-to-rank \citep{Joachims2002svm}. The standardization operation reduces the internal covariate shift \citep{Ioffe15bn}. That is, it avoids the complications introduced by changes in the distribution of document feature vectors during learning. In addition, as the document feature vectors are learned from scratch as well, the standardization forces the learned feature vectors to be centered around the null vector. Afterwards, we apply the hard-saturating nonlinearity $\HardTanH{}$ \citep{Gulcehre2016noisy} such that the feature activations are between $-1$ and $1$.

The objective is to maximize the similarity between
\begin{equation*}
\NGramProjection{}\text{ and }\vec{R}_\Documents^{(B_i^{(\Document)})}, 
\end{equation*}
while minimizing the similarity between 
\[
\NGramProjection{}
\] 
and the representations of other documents. Therefore, a document is characterized by the concepts it contains, and consequently, documents describing similar concepts will cluster together, as postulated by the \textbf{clustering hypothesis} \citep{Rijsbergen1979}. \ModelName{} strongly relies on the clustering hypothesis as ranking is performed according to nearest neighbor retrieval (Eq.~\ref{eq:score}).

Considering the full set of documents $\Documents{}$ is often costly as $\Length{\Documents{}}$ can be large. Therefore, we apply an adjusted-for-bias variant of negative sampling \citep{VanGysel2016products,Mikolov2013compositionality,Gutmann2010}, where we uniformly sample negative examples from $\Documents{}$. Adopted from \citep{VanGysel2016products}, we define
\newcommand{\SimilarityRandomVariable}{\mathcal{S}}
\newcommand{\SigmoidFn}{\sigma}
\newcommand{\ExpFn}{\text{exp}}
\newcommand{\ProbSimilarEntity}[1]{\CondProb{\SimilarityRandomVariable{}}{#1, \BatchInstancePhrase{i}}}
\newcommand{\ProbMassEntity}[1]{\Apply{\SigmoidFn{}}{\DotProduct{#1}{\NGramProjection{}}}}
\newcommand{\NumNegativeExamples}{z}
\begin{equation}
\ProbSimilarEntity{\Document{}} = 
\sigma\left(\vec{R}_\Documents^{(\Document)}\cdot \NGramProjection{} \right)
%\ProbMassEntity{\DocumentEmbedding{\Document{}}}
\end{equation}
as the similarity of two representations in latent vector space, where
\begin{equation*}
\Apply{\SigmoidFn{}}{t} = \frac{1}{1 + \Apply{\ExpFn{}}{-t}}
\end{equation*}
denotes the sigmoid function and $\SimilarityRandomVariable{}$ is an indicator binary random variable that says whether the representation of document $\Document{}$ is similar to the projection of n-gram $\BatchInstancePhrase{i}$.

The probability of document $\BatchInstanceDocument{i}$ given phrase $\BatchInstancePhrase{i}$ is then approximated by uniformly sampling $\NumNegativeExamples{}$ contrastive examples:
\begin{eqnarray}
\label{eq:instance_loss}
\lefteqn{\log\CondApproxProb{\BatchInstanceDocument{i}}{\BatchInstancePhrase{i}} = \qquad\qquad} \\
& & \frac{\NumNegativeExamples{} + 1}{2 \NumNegativeExamples{}} \Bigg( \NumNegativeExamples{} \log\ProbSimilarEntity{\BatchInstanceDocument{i}} + \nonumber\\
& & \sum^{\NumNegativeExamples{}}_{\substack{k=1, \\ \Document{}_k \sim U(\Documents{})}} \log \left( 1.0 - \ProbSimilarEntity{\Document{}_k} \right) \Bigg), \nonumber
\end{eqnarray}
where $U(\Documents{})$ denotes the uniform distribution over documents $\Documents{}$, the distribution used for obtaining negative examples \citep{VanGysel2016products,Gutmann2010}. Then, the \textbf{loss function} we use to optimize our model is Eq.~\ref{eq:instance_loss} averaged over the instances in batch $\Batch{}$:
\begin{eqnarray}
\lefteqn{\Apply{\LossFn{}}{\WordMatrix{}, \DocumentMatrix{}, \TransformMatrix{}, \ErrorBias{} \mid \Batch{}}} \\
& = & - \frac{1}{\BatchSize{}} \sum_{i=1}^{\BatchSize{}} \log\CondApproxProb{\BatchInstanceDocument{i}}{\BatchInstancePhrase{i}} \nonumber \\ \nonumber
& &\quad{}+ \frac{\lambda}{2 \BatchSize{}} \left( \sum_{i,j} \WordMatrix{}^2_{i,j} + \sum_{i,j} \DocumentMatrix{}^2_{i,j} + \sum_{i,j} \TransformMatrix{}^2_{i,j} \right),
\end{eqnarray}
where $\lambda$ is a weight regularization hyper-parameter. We optimize our parameters $\theta{}$ ($\WordMatrix{}$, $\DocumentMatrix{}$, $\TransformMatrix{}$ and $\ErrorBias{}$) using Adam~\citep{Kingma2014}, a first-order gradient-based optimization function for stochastic objective functions that is very similar to momentum. The update rule for parameter $\theta{}$ given a batch $\Batch{}$ at batch update time $t$ equals:
\newcommand{\FirstMoment}[1]{\hat{m}_{#1}^{(t)}}
\newcommand{\SecondMoment}[1]{\hat{v}_{#1}^{(t)}}
\begin{equation}
\label{eq:adam}
\theta{}^{(t + 1)} \leftarrow \theta{}^{(t)} - \alpha \frac{\FirstMoment{\theta{}}}{\sqrt{\SecondMoment{\theta{}}} + \epsilon},
\end{equation}
where $\FirstMoment{\theta{}}$ and $\SecondMoment{\theta{}}$, respectively, are the first and second moment estimates (over batch update times) \citep{Kingma2014} of the gradient of the loss $\Apply{\frac{\partial \LossFn}{\partial \theta{}}}{\WordMatrix{}, \DocumentMatrix{}, \TransformMatrix{}, \ErrorBias{} \mid \Batch{}}$ w.r.t.\ parameter $\theta{}$ at batch update time $t$ and $\epsilon = 10^{-8}$ is a constant to ensure numerical stability. The use of this optimization method causes every parameter to be updated with every batch, unlike regular stochastic gradient descent, where the only parameters that are updated are those for which there is a non-zero gradient of the loss. This is important in \ModelName{} due to the large number of word and document vectors.

The algorithmic contributions of \ModelName{} over \LSE{} are the components of the objective mentioned next. Eq.~\ref{eq:raw_projection} forces individual words to compete in order to contribute to the resulting n-gram representation. Consequently, non-discriminative words will have a small L2-norm. In Eq.~\ref{eq:projection} we perform standardization to reduce the internal covariate shift \citep{Ioffe15bn}. In addition, the standardization forces n-gram representations to distinguish themselves only in the dimensions that matter for matching. Frequent words are naturally prevalent in n-grams, however, they have low discriminative power as they are non-specific. The bias $\ErrorBias{}$ captures word frequency regularities that are non-discriminative for the semantic concepts within the respective n-gram/document pairs and allows the transformation in Eq.~\ref{eq:transform} to focus on concept matching. The re-weighting of the positive instance in Eq.~\ref{eq:instance_loss} removes a dependence on the number of negative examples $\NumNegativeExamples{}$ where a large $\NumNegativeExamples{}$ presented the model with a bias towards negative examples.

\subsection{Implementation}
\label{sec:impl}

The major cause of technical challenges of the \ModelName{} training procedure is not due to time complexity, but rather space restrictions. This is because we mitigate expensive computation by estimating vector space models using graphics processing units (GPUs). The main limitation of these massively-parallel computation devices is that they rely on their own memory units. Consequently, parameters and intermediate results of the training procedure need to persist in the GPU memory space. The asymptotic space complexity of the parameters equals:
\begin{equation*}
\Apply{O}{\underbrace{\Length{\Vocabulary{}} \times \LatentWordDim{}}_{\text{word representations } \WordMatrix{}} + \underbrace{\LatentDocumentDim{} \times \LatentWordDim{}}_{\text{transform } \TransformMatrix{}} + \underbrace{\Length{\Documents{}} \times \LatentDocumentDim{}}_{\text{document representations } \DocumentMatrix{}}}.
\end{equation*}
In addition, Eq.~\ref{eq:adam} requires us to keep the first and second moment of the gradient over time for every parameter in memory. Therefore, for every parameter, we retain three floating point values in memory at all times.
%
% 1 000 000 * 256 + 256 * (300 + 1) + 64 000 * 300 = 275 277 056
% 275 277 056 * 3 * 4 bytes = 3.30332467 gigabytes
%
For example, if we have a collection of \numprint{1}M documents (256-dim.)\ with a vocabulary of \numprint{64}K terms (300-dim.), then the model has ${\sim}$\numprint{275}M parameters. Consequently, under the assumption of 32-bit floating point, the resident memory set has a best-case least upper bound of \numprint{3.30}GB memory. The scalability of our method---similar to all latent vector space models---is determined by the number of documents within the collection. However, the current generation of GPUs---that typically boast around \numprint{12}GB of memory---can comfortably be used to train models of collections consisting of up to 2 million documents. In fact, next-generation GPUs have double the amount of memory---\numprint{24}GB---and this amount will likely increase with the introduction of future processing units \citep{Wiki2017nvidia}. This, and the development of distributed GPU technology \citep{Tensorflow2015whitepaper}, leads us to believe that the applicability of our approach to larger retrieval domains is simply a matter of time \citep{Moore1998cramming}.

In addition to the scarcity of memory on current generation GPUs, operations such as the averaging of word to n-gram representations (Eq.~\ref{eq:averaging}) can be performed in-place. However, these critical optimizations are not available in general-purpose machine learning toolkits. Therefore, we have implemented the \ModelName{} training procedure directly in C++/CUDA, such that we can make efficient use of sparseness and avoid unnecessary memory usage. In addition, models are stored in the open HDF5 format \citep{Folk2011hdf5} and the toolkit provides a Python module that can be used to query trained \ModelName{} models on the CPU. This way, a trained \ModelName{} can easily be integrated in existing applications. The toolkit is licensed under the permissive MIT open-source license; see footnote~\ref{fn:ImplementationURL}.

% !TEX root = ./tois2018-adhoc-retrieval.tex

\section{Experimental Setup}
\label{sec:experimental}

\subsection{Research questions}

\newcommand{\RQ}[2]{
    \begin{description}
    \phantomsection\label{section:setup:rq#1}
    \item[RQ#1] #2
    \end{description}
}

\newcommand{\RQRef}[1]{\textbf{\hyperref[section:setup:rq#1]{RQ#1}}}

In this paper we investigate the viability of neural representation learning methods for semantic matching in document search. We seek to answer the following research questions:

\RQ{1}{\ResearchQuestionOne{}}

\noindent%
In particular, how does it compare to other methods that represent queries/documents as low-dimensional vectors? What is the difference in performance with purely lexical models that perform exact term matching and represent queries/documents as bag-of-words vectors?

\RQ{2}{\ResearchQuestionTwo{}}

\noindent%
Does \ModelName{} improve over other retrieval models only on a handful of queries? Instead of computing averages over queries, what if we look at the pairwise differences between rankers?

\RQ{3}{\ResearchQuestionThree{}}

\noindent%
Can we use the differences that we observe in \RQRef{2} between \ModelName{}, \QLM{} and other latent vector space models to our advantage to improve retrieval performance? Can we pinpoint where the improvements of \ModelName{} come from?

\RQ{4}{\ResearchQuestionFour{}}

\noindent%
Traditional retrieval models such as generative language models are known to incorporate corpus statistics regarding term specificity \citep{Sparck1972specificity,Robertson2004idf} and document length \citep{Zhai2004smoothing}. Are similar corpus statistics present in \ModelName{} and what does this tell us about the ranking task?

\subsection{Benchmark datasets \& experiments}
\label{sec:benchmarks}

In this paper we are interested in query/document matching. Therefore, we evaluate \ModelName{} on newswire article collections from TREC. Other retrieval domains, such as web search or social media, deal with aspects such as document freshness/importance and social/hyperlink/click graphs that may obfuscate the impact of matching queries to documents.
Therefore, we follow the experimental setup of \citet{Zhai2004smoothing} and use four article retrieval sub-collections from the TIPSTER corpus \citep{Harman1992tipster}: \BenchmarkAP{} (\ShortBenchmarkAP{}), \BenchmarkFT{} (\ShortBenchmarkFT{}), \BenchmarkLATimes{} (\ShortBenchmarkLATimes{}) and \BenchmarkWSJ{} (\ShortBenchmarkWSJ{}) \citep{Harman1993document}. In addition, we consider the \BenchmarkRobust{} collection that constitutes of Disk 4/5 of the TIPSTER corpus without the Congressional Record and the \BenchmarkNYT{} collection that consists of articles written and published by the New York Times between 1987 and 2007. For evaluation, we take topics \numprint{50}--\numprint{200} from TREC 1--3\footnote{We only consider judgments corresponding to each of the sub-collections.\label{footnote:subcollections}} (\ShortBenchmarkAP{}, \ShortBenchmarkWSJ{}), topics \numprint{301}--\numprint{450} from TREC 6--8\footnotemark[\thefootnote]{} (\ShortBenchmarkFT{}, \ShortBenchmarkLATimes{}) \cite{TRECAdhoc}, topics \numprint{301}--\numprint{450}, \numprint{601}--\numprint{700} from \BenchmarkRobust{} \citep{Voorhees2005robust} and the 50 topics assessed by NIST (a subset of the \BenchmarkRobust{} topics judged for the \ShortBenchmarkNYT{} collection) during the TREC 2017 Common Core Track \citep{Allan2017treccommon}. From each topic we take its title as the corresponding query. Topics without relevant documents are filtered out. We randomly create a split\footnote{The validation/test splits can be found at \paperImplementationUrl{}.} of validation (20\%) and test (80\%) queries (with the exception of the \ShortBenchmarkNYT{} collection). For the \ShortBenchmarkNYT{} collection, we select the hyperparameter configuration that optimizes the \BenchmarkRobust{} validation set on the \BenchmarkRobust{} collection and we take the 50 queries assessed by NIST and their judgments---specifically created for the \ShortBenchmarkNYT{} collection---as our test set. This way, method hyperparameters are optimized on the validation set (as described in Section~\ref{sec:retrievalmodels}) and the retrieval performance is reported on the test set; see Table~\ref{tbl:adhoc_stats}.

The inclusion of early TREC collections (\ShortBenchmarkAP{}, \ShortBenchmarkFT{}, \ShortBenchmarkLATimes{} and \ShortBenchmarkWSJ{}) is motivated by the fact that during the first few years of TREC, there was a big emphasis on submissions where the query was constructed manually from each topic and interactive feedback was used \citep{Harman1993document}. That is, domain experts repeatedly formulated manual queries using the full topic (title, description, narrative), observed the obtained rankings and then reformulated their query in order to obtain better rankings. Consequently, these test collections are very useful when evaluating semantic matches as relevant documents do not necessarily contain topic title terms. From TREC-5 and onwards, less emphasis was put on rankings generated using interactive feedback and shifted towards automated systems only \citep[footnote 1]{Harman1996trec5}. In fact, within the \BenchmarkRobust{} track, only automated systems were submitted \citep[\S2]{Voorhees2005robust} due to the large number of (a) documents ($\sim$500K) and (b) topics (250). In the 2017 Common Core Track \citep{Allan2017treccommon}, interactive rankings were once again submitted by participants, in addition to rankings obtained by the latent vector space model presented in this paper.

\newcommand{\Domain}[3]{
\subsubsection{#2}
\label{sec:#1}
#3}

\newcommand{\DomainStatsTable}[1]{
\begin{table*}[t]
\centering

\caption{Overview of the retrieval benchmarks. T and V denote the test and validation sets, respectively. Arithmetic mean and standard deviation are reported wherever applicable.\label{tbl:#1_stats}}

\scalebox{1.0}{%
\centering%
\begin{tabular}{c}%
\input{resources/#1_stats.tex}%
\end{tabular}}
\end{table*}%
}

\newcommand{\CF}[1]{\text{CF}_{#1}}
\newcommand{\TF}[1]{\text{TF}_{#1}}
\newcommand{\IDF}[1]{\text{IDF}_{#1}}
{
\renewcommand{\BenchmarkAP}{\ShortBenchmarkAP}
\renewcommand{\BenchmarkWSJ}{\ShortBenchmarkWSJ}
\renewcommand{\BenchmarkLATimes}{\ShortBenchmarkLATimes}
\renewcommand{\BenchmarkFT}{\ShortBenchmarkFT}
\renewcommand{\BenchmarkNYT}{\ShortBenchmarkNYT}

\begin{table*}[t]
\centering

\caption{Overview of the retrieval benchmarks. T and V denote the test and validation sets, respectively. Arithmetic mean and standard deviation are reported wherever applicable.\label{tbl:adhoc_stats}}

\scalebox{1.0}{%
\centering%
\begin{tabular}{c}%
\begin{tabular}{lccc}%
\toprule%
& \BenchmarkAP{} & \BenchmarkFT{} & \BenchmarkLATimes{} \\%
\midrule%
\textbf{Collection} (training) \\%
Documents & \phantom{}\numprint{164597}  & \phantom{}\numprint{210158}  & \phantom{}\numprint{131896} \\%
Document length & {\renewcommand{\arraystretch}{0.8}\setlength{\tabcolsep}{8pt}\begin{tabular}[t]{ll}&\phantom{}\nprounddigits{2}\npdecimalsign{.}\numprint{461.6340698797704}$\,\pm\,$\phantom{}\nprounddigits{2}\npdecimalsign{.}\numprint{243.61286335206674}  \\ \end{tabular}} & {\renewcommand{\arraystretch}{0.8}\setlength{\tabcolsep}{8pt}\begin{tabular}[t]{ll}&\phantom{}\nprounddigits{2}\npdecimalsign{.}\numprint{399.6802120309481}$\,\pm\,$\phantom{}\nprounddigits{2}\npdecimalsign{.}\numprint{366.4147351756061}  \\ \end{tabular}} & {\renewcommand{\arraystretch}{0.8}\setlength{\tabcolsep}{8pt}\begin{tabular}[t]{ll}&\phantom{}\nprounddigits{2}\npdecimalsign{.}\numprint{502.1807333050321}$\,\pm\,$\phantom{}\nprounddigits{2}\npdecimalsign{.}\numprint{519.5806529063212}  \\ \end{tabular}}\\%
Unique terms & \nprounddigits{2}\npdecimalsign{.}\numprint{\xintFloat [3]{267318}}  & \nprounddigits{2}\npdecimalsign{.}\numprint{\xintFloat [3]{305310}}  & \nprounddigits{2}\npdecimalsign{.}\numprint{\xintFloat [3]{267156}} \\%
\textbf{Queries} (testing) & {\renewcommand{\arraystretch}{1.0}\begin{tabular}[t]{ll} &(T) \phantom{}\numprint{119}  \\  &(V)\phantom{0}\numprint{30}  \\ \end{tabular}} & {\renewcommand{\arraystretch}{1.0}\begin{tabular}[t]{ll} &(T) \phantom{}\numprint{116}  \\  &(V)\phantom{0}\numprint{28}  \\ \end{tabular}} & {\renewcommand{\arraystretch}{1.0}\begin{tabular}[t]{ll} &(T) \phantom{}\numprint{113}  \\  &(V)\phantom{0}\numprint{30}  \\ \end{tabular}}\\%
Query terms & {\renewcommand{\arraystretch}{0.8}\setlength{\tabcolsep}{8pt}\begin{tabular}[t]{ll}\phantom{(T)}&\phantom{00}\nprounddigits{2}\npdecimalsign{.}\numprint{5.06}$\,\pm\,$\phantom{00}\nprounddigits{2}\npdecimalsign{.}\numprint{3.14}  \\ \end{tabular}} & {\renewcommand{\arraystretch}{0.8}\setlength{\tabcolsep}{8pt}\begin{tabular}[t]{ll}\phantom{(T)}&\phantom{00}\nprounddigits{2}\npdecimalsign{.}\numprint{2.50}$\,\pm\,$\phantom{00}\nprounddigits{2}\npdecimalsign{.}\numprint{0.69}  \\ \end{tabular}} & {\renewcommand{\arraystretch}{0.8}\setlength{\tabcolsep}{8pt}\begin{tabular}[t]{ll}\phantom{(T)}&\phantom{00}\nprounddigits{2}\npdecimalsign{.}\numprint{2.48}$\,\pm\,$\phantom{00}\nprounddigits{2}\npdecimalsign{.}\numprint{0.69}  \\ \end{tabular}}\\%
Relevant documents & {\renewcommand{\arraystretch}{0.8}\setlength{\tabcolsep}{8pt}\begin{tabular}[t]{ll}(T)&\phantom{}\nprounddigits{2}\npdecimalsign{.}\numprint{111.45}$\,\pm\,$\phantom{}\nprounddigits{2}\npdecimalsign{.}\numprint{136.01}  \\ (V)&\phantom{0}\nprounddigits{2}\npdecimalsign{.}\numprint{86.43}$\,\pm\,$\phantom{0}\nprounddigits{2}\npdecimalsign{.}\numprint{72.63}  \\ \end{tabular}} & {\renewcommand{\arraystretch}{0.8}\setlength{\tabcolsep}{8pt}\begin{tabular}[t]{ll}(T)&\phantom{0}\nprounddigits{2}\npdecimalsign{.}\numprint{34.91}$\,\pm\,$\phantom{0}\nprounddigits{2}\npdecimalsign{.}\numprint{42.94}  \\ (V)&\phantom{0}\nprounddigits{2}\npdecimalsign{.}\numprint{30.46}$\,\pm\,$\phantom{0}\nprounddigits{2}\npdecimalsign{.}\numprint{26.97}  \\ \end{tabular}} & {\renewcommand{\arraystretch}{0.8}\setlength{\tabcolsep}{8pt}\begin{tabular}[t]{ll}(T)&\phantom{0}\nprounddigits{2}\npdecimalsign{.}\numprint{24.83}$\,\pm\,$\phantom{0}\nprounddigits{2}\npdecimalsign{.}\numprint{34.31}  \\ (V)&\phantom{0}\nprounddigits{2}\npdecimalsign{.}\numprint{24.30}$\,\pm\,$\phantom{0}\nprounddigits{2}\npdecimalsign{.}\numprint{21.19}  \\ \end{tabular}}\\%
\\%
\midrule%
\end{tabular}%
\\%
\begin{tabular}{lccc}%
& \BenchmarkNYT{} & \BenchmarkRobust{} & \BenchmarkWSJ{} \\%
\midrule%
\textbf{Collection} (training) \\%
Documents & \phantom{}\numprint{1855658}  & \phantom{}\numprint{528155}  & \phantom{}\numprint{173252} \\%
Document length & {\renewcommand{\arraystretch}{0.8}\setlength{\tabcolsep}{8pt}\begin{tabular}[t]{ll}&\phantom{}\nprounddigits{2}\npdecimalsign{.}\numprint{572.1752801432273}$\,\pm\,$\phantom{}\nprounddigits{2}\npdecimalsign{.}\numprint{605.8240047996469}  \\ \end{tabular}} & {\renewcommand{\arraystretch}{0.8}\setlength{\tabcolsep}{8pt}\begin{tabular}[t]{ll}&\phantom{}\nprounddigits{2}\npdecimalsign{.}\numprint{479.7217653908449}$\,\pm\,$\phantom{}\nprounddigits{2}\npdecimalsign{.}\numprint{869.2695174032617}  \\ \end{tabular}} & {\renewcommand{\arraystretch}{0.8}\setlength{\tabcolsep}{8pt}\begin{tabular}[t]{ll}&\phantom{}\nprounddigits{2}\npdecimalsign{.}\numprint{447.5146491815376}$\,\pm\,$\phantom{}\nprounddigits{2}\npdecimalsign{.}\numprint{454.7468232217638}  \\ \end{tabular}}\\%
Unique terms & \nprounddigits{2}\npdecimalsign{.}\numprint{\xintFloat [3]{1351413}}  & \nprounddigits{2}\npdecimalsign{.}\numprint{\xintFloat [3]{782799}}  & \nprounddigits{2}\npdecimalsign{.}\numprint{\xintFloat [3]{250164}} \\%
\textbf{Queries} (testing) & {\renewcommand{\arraystretch}{1.0}\begin{tabular}[t]{ll} &(T) \phantom{0}\numprint{50}  \\  & \\ \end{tabular}} & {\renewcommand{\arraystretch}{1.0}\begin{tabular}[t]{ll} &(T) \phantom{}\numprint{200}  \\  &(V)\phantom{0}\numprint{49}  \\ \end{tabular}} & {\renewcommand{\arraystretch}{1.0}\begin{tabular}[t]{ll} &(T) \phantom{}\numprint{120}  \\  &(V)\phantom{0}\numprint{30}  \\ \end{tabular}}\\%
Query terms & {\renewcommand{\arraystretch}{0.8}\setlength{\tabcolsep}{8pt}\begin{tabular}[t]{ll}\phantom{(T)}&\phantom{00}\nprounddigits{2}\npdecimalsign{.}\numprint{6.58}$\,\pm\,$\phantom{00}\nprounddigits{2}\npdecimalsign{.}\numprint{0.70}  \\ \end{tabular}} & {\renewcommand{\arraystretch}{0.8}\setlength{\tabcolsep}{8pt}\begin{tabular}[t]{ll}\phantom{(T)}&\phantom{00}\nprounddigits{2}\npdecimalsign{.}\numprint{5.28}$\,\pm\,$\phantom{00}\nprounddigits{2}\npdecimalsign{.}\numprint{0.74}  \\ \end{tabular}} & {\renewcommand{\arraystretch}{0.8}\setlength{\tabcolsep}{8pt}\begin{tabular}[t]{ll}\phantom{(T)}&\phantom{00}\nprounddigits{2}\npdecimalsign{.}\numprint{5.05}$\,\pm\,$\phantom{00}\nprounddigits{2}\npdecimalsign{.}\numprint{3.14}  \\ \end{tabular}}\\%
Relevant documents & {\renewcommand{\arraystretch}{0.8}\setlength{\tabcolsep}{8pt}\begin{tabular}[t]{ll}(T)&\phantom{}\nprounddigits{2}\npdecimalsign{.}\numprint{180.04}$\,\pm\,$\phantom{}\nprounddigits{2}\npdecimalsign{.}\numprint{132.74}  \\  \\ \end{tabular}} & {\renewcommand{\arraystretch}{0.8}\setlength{\tabcolsep}{8pt}\begin{tabular}[t]{ll}(T)&\phantom{0}\nprounddigits{2}\npdecimalsign{.}\numprint{70.33}$\,\pm\,$\phantom{0}\nprounddigits{2}\npdecimalsign{.}\numprint{73.68}  \\ (V)&\phantom{0}\nprounddigits{2}\npdecimalsign{.}\numprint{68.27}$\,\pm\,$\phantom{0}\nprounddigits{2}\npdecimalsign{.}\numprint{77.41}  \\ \end{tabular}} & {\renewcommand{\arraystretch}{0.8}\setlength{\tabcolsep}{8pt}\begin{tabular}[t]{ll}(T)&\phantom{0}\nprounddigits{2}\npdecimalsign{.}\numprint{96.99}$\,\pm\,$\phantom{0}\nprounddigits{2}\npdecimalsign{.}\numprint{93.30}  \\ (V)&\phantom{}\nprounddigits{2}\npdecimalsign{.}\numprint{101.93}$\,\pm\,$\phantom{}\nprounddigits{2}\npdecimalsign{.}\numprint{117.65}  \\ \end{tabular}}\\%
\\%
\bottomrule%
\end{tabular}%
\\%
\end{tabular}}
\end{table*}%

}

We address \RQRef{1} by comparing \ModelName{} to latent retrieval models (detailed in Section~\ref{sec:retrievalmodels}). In addition, we perform a per-query pairwise comparison of methods where we look at what method performs best for each query in terms of \MAPCut{} (\RQRef{2}). A method performs better or worse than another if the absolute difference in \MAPCut{} exceeds $\delta = 0.01$; otherwise, the two methods perform similar. To address \RQRef{3}, we consider the combinations (Section~\ref{sec:ltr}) of \QLM{} with \ModelName{} and the strongest latent vector space baseline of \RQRef{1}. That is, \FullWordToVec{} where the summands are weighted using self-information. In addition, we look at the correlation between per-query \TitlestatRelevant{} (see Section~\ref{sec:evaluationmeasures}) and the pairwise differences in \MAPCut{} between \ModelName{} and all the other retrieval models. A positive correlation indicates that \ModelName{} is better at lexical matching than the other method, and vice versa for a negative correlation. For \RQRef{4}, we examine the relation between the collection frequency $\CF{\Word{}}$ and the L2-norm of their word embeddings $\Apply{\CompositionalFn{}}{\Word{}}$ for all terms $\Word{} \in \Vocabulary{}$.

\subsection{Retrieval models considered for comparison}
\label{sec:retrievalmodels}

The document collection is first indexed by Indri\footnote{Stopwords are removed using the standard stopword list of Indri.} \citep{Strohman2005indri}. Retrieval models not implemented by Indri access the underlying tokenized document collection using \texttt{pyndri} \citep{VanGysel2017pyndri}. This way, all methods compared in this paper parse the text collection consistently. 

\subsubsection{Models compared}

The key focus of this paper is the alleviation of the vocabulary gap in information retrieval and consequently, in theory, we score all documents in each collection for every query. In practice, however, we rely on nearest neighbor search algorithms to retrieve the top-k documents \citep{Boytsov2016knn}. Note that this is in contrast to many other semantic matching methods \citep{Zuccon2015nntm,Ai2016doc2veclm,Nalisnick2016desm} that have only been shown to perform well in document re-ranking scenarios where an initial pool of candidate documents is retrieved using a lexical matching method. However, candidate documents most affected by the vocabulary gap (i.e., relevant documents that do not contain any query terms) will simply remain undiscovered in a re-ranking scenario and consequently we compare \ModelName{} only to latent vector space models that can be queried using a nearest neighbor search.

The following latent vector space models are compared:
\begin{enumerate}
	\item \FullDocToVec{} (\DocToVec{}) \citep{Le2014} with the distributed memory architecture. The pre-processing of document texts to learn latent document representations is a topic of study by itself and its effects are outside the scope of this work. Consequently, we disable vocabulary filtering and frequent word subsampling in order to keep the input to all representation learning algorithms consistent. We sweep the one-sided window size and the embedding size respectively in partitions $\{x / 2 \mid x = 4, 6, 8, 10, 12, 16, 24, 32\}$ and $\{64, 128, 256\}$ on the validation set. Models are trained for \numprint{15} iterations and we select the model iteration that performs best on the validation set. Documents are ranked in decreasing order of the cosine similarity between the document representation and the average of the word embeddings in the query.
	\item \FullWordToVec{} (\WordToVec{}) \citep{Mikolov2013word2vec,Vulic2015monolingual} with the Skip-Gram architecture. We follow the method introduced by \citet{Vulic2015monolingual} where query/document representations are constructed by composing the representations of the words contained within them. We consider both the unweighted sum (\CombineAdd{}) and the sum of vectors weighted by the term's self-information (\CombineSelfInformation{}). Self-information is a term specificity measure similar to Inverse Document Frequency (IDF) \citep{Cover2012informationtheory}. The hyperparameters of \FullWordToVec{} are swept in the same manner as \FullDocToVec{}.
	\item Latent Semantic Indexing (\LSI{}) \citep{Deerwester1990} with TF-IDF weighting and the number of topics $K \in \{64, 128, 256\}$ optimized on the validation set.
	\item Latent Dirichlet Allocation (\LDA{}) \citep{Blei2003} with $\alpha = \beta = 0.1$ and the number of topics $K \in \{64, 128, 256\}$ optimized on the validation set. We train the model for 100 iterations or until topic convergence is achieved. Documents are ranked in decreasing order of the cosine similarity between the query topic distribution and the document topic distribution.
	\item Representation learning methods \LSE{} \citep{VanGysel2016products} and \ModelName{} (this paper). For hyperparameters, we largely follow the findings of \citep{VanGysel2016products}: word representation dim. $\LatentWordDim{} = 300$, number of negative examples $\NumNegativeExamples{} = 10$, learning rate $\alpha = 0.001$, regularization lambda $\lambda = 0.01$. For \LSE{}, batch size $\BatchSize{} = 4096$ (as in \citep{VanGysel2016products}), while for \ModelName{} the batch size $\BatchSize{} = 51200$ (empirically determined on a holdout document collection that we did not include in this paper). The dimensionality of the document representations $\LatentDocumentDim{} \in \{64, 128, 256\}$ and the n-gram size $\NGramSize{} \in \{4, 6, 8, 10, 12, 16, 24, 32\}$ are optimized on the validation set. Similar to \DocToVec{}, models are trained for \numprint{15} iterations on the training set and we select the model iteration that performs best on the validation set; a single iteration consists of $\ceil{\frac{1}{\BatchSize{}} \sum_{\Document{} \in \Documents{}} \left(\Length{\Document{}} - \NGramSize{} + 1 \right)}$ batches.
\end{enumerate}

In addition, we consider lexical language models (\QLM{}) \citep{Zhai2004smoothing} using the Indri engine with both Dirichlet (\Dirichlet{}) and Jelinek-Mercer (\JelinekMercer{}) smoothing; smoothing hyperparameters $\mu \in \splitatcommas{\{125, 250, 500, 750, 1000, 2000, 3000, 4000, 5000\}}$ and $\lambda \in \{ x \mid k \in \mathbb{N}_{>0}, k \leq 20, x = {k}/{20}\}$, respectively, are optimized on the validation set. The retrieval effectiveness of \QLM{} is provided as a point of reference in \RQRef{1}. For \RQRef{2}, the \QLM{} is used as a lexical retrieval model that is fused with latent vector space models to provide a mixture of lexical and semantic matching.

For the latent vector spaces (\DocToVec{}, \LSI{}, \LDA{}, \LSE{} and \ModelName{}), the vocabulary size is limited to the top-60k most frequent words as per \citep{VanGysel2016products}, given that latent methods rely on word co-occurrence in order to learn latent relations. For \FullDocToVec{}, \FullWordToVec{}, \LSI{} and \LDA{} we use the Gensim\footnote{\url{https://github.com/RaRe-Technologies/gensim}} implementation; the neural representation learning methods use our open source CUDA implementation described in Section~\ref{sec:impl}; again, see footnote~\ref{fn:ImplementationURL}.

\subsubsection{Combinations of \QLM{} and latent features}
\label{sec:ltr}

We combine individual rankers by performing a grid search on the weights of a linear combination using 20-fold cross validation on the test sets (Table~\ref{tbl:adhoc_stats}). For \QLM{}, feature values are the log-probabilities of the query given the document, while for the latent features (\ModelName{} and \WordToVecSgSi{}), we use the cosine similarity between query/document representations. For every feature weight, we sweep between $0.0$ and $1.0$ with increments of $0.0125$ on the fold training set. Individual features are normalized per query such that their values lie between $0$ and $1$. We select the weight configuration that achieves highest Mean Average Precision on the training set and use that configuration to score the test set. During scoring of the fold test set, we take the pool of the top-1k documents ranked by the individual features as candidate set.

\subsection{Evaluation measures and statistical significance}
\label{sec:evaluationmeasures}

\newcommand{\CorrelationSignificant}{^{\dagger}}

\newcommand{\Significant}{^{*\phantom{**}}}
\newcommand{\MoreSignificant}{^{**\phantom{*}}}
\newcommand{\HighlySignificant}{^{***}}
\newcommand{\UnknownSignificant}{\phantom{\HighlySignificant{}}}

To address \RQRef{1}, \RQRef{2} and \RQRef{3}, we report Mean Average Precision at rank 1000 (\MAPCut{}\footnote{\MAPCut{} summarizes the precision/recall curve and thus represents the trade-off between precision and recall at different rank cut-offs. When using latent vector space models for generating a set of candidate documents (e.g., for subsequent re-ranking), the precision/recall trade-off translates into precision for a particular candidate set size. A smaller candidate set is preferred as re-ranking many candidate documents can be expensive.}), Normalized Discounted Cumulative Gain at rank 100 (\NDCGCut{}) and Precision at rank 10 (\Precision{}) to measure retrieval effectiveness. For \RQRef{3}, we also look at the per-query \TitlestatRelevant{} \citep{Buckley2007bias}, the expected normalized term overlap between query and document. All evaluation measures are computed using TREC's official evaluation tool, {\tt trec\_eval}.\footnote{\url{https://github.com/usnistgov/trec_eval}} Wherever reported, significance of observed differences is determined using a two-tailed paired Student's t-test \citep{Smucker2007} ($\HighlySignificant{} \, p < 0.01$;  $\MoreSignificant{} \, p < 0.05$; $\Significant{} \, p < 0.1$). For correlation coefficients, significance is determined using a permutation test ($\CorrelationSignificant{} \, p < 0.01$). For \RQRef{4}, we use Welch's t-test to determine whether the mean L2-norm of mid-frequency (middle-50\%) words is significantly different from the mean L2-norm of low- (bottom 25\%) and high-frequency (top 25\%) words.

% !TEX root = ./tois2018-adhoc-retrieval.tex

\newcommand{\RQAnswer}[3]{
\begin{description}
	\item[\RQRef{#1}] #2
\end{description}
\noindent #3
}

\newcommand{\Table}[4][]{
\begin{#2}
\centering
\caption{#1\label{tbl:#3_results}}
\scalebox{#4}{%
\renewcommand{\MAPCut}{MAP}%
\renewcommand{\NDCGCut}{NDCG}%
\renewcommand{\BenchmarkAP}{\ShortBenchmarkAP}%
\renewcommand{\BenchmarkWSJ}{\ShortBenchmarkWSJ}%
\renewcommand{\BenchmarkLATimes}{\ShortBenchmarkLATimes}%
\renewcommand{\BenchmarkFT}{\ShortBenchmarkFT}%
\renewcommand{\BenchmarkNYT}{\ShortBenchmarkNYT}%
\centering%
\begin{tabular}{c}%
\input{resources/#3.tex}%
\end{tabular}%
}%
\end{#2}%
}

\newcommand{\DoubleColumnTable}[2][]{

\begin{table*}
\centering
\caption{#1\label{tbl:#2_results}}
\scalebox{1.0}{%
\renewcommand{\MAPCut}{MAP}%
\renewcommand{\NDCGCut}{NDCG}%
\renewcommand{\BenchmarkAP}{\ShortBenchmarkAP}%
\renewcommand{\BenchmarkWSJ}{\ShortBenchmarkWSJ}%
\renewcommand{\BenchmarkLATimes}{\ShortBenchmarkLATimes}%
\renewcommand{\BenchmarkFT}{\ShortBenchmarkFT}%
\renewcommand{\BenchmarkNYT}{\ShortBenchmarkNYT}%
\centering%
\begin{tabular}{c}%
\input{resources/#2.tex}%
\end{tabular}%
}%
\end{table*}%

}

\newcommand{\SingleColumnTable}[2][]{

\begin{table}
\centering
\caption{#1\label{tbl:#2_results}}
\scalebox{1.0}{%
\renewcommand{\MAPCut}{MAP}%
\renewcommand{\NDCGCut}{NDCG}%
\renewcommand{\BenchmarkAP}{\ShortBenchmarkAP}%
\renewcommand{\BenchmarkWSJ}{\ShortBenchmarkWSJ}%
\renewcommand{\BenchmarkLATimes}{\ShortBenchmarkLATimes}%
\renewcommand{\BenchmarkFT}{\ShortBenchmarkFT}%
\renewcommand{\BenchmarkNYT}{\ShortBenchmarkNYT}%
\centering%
\begin{tabular}{c}%
\input{resources/#2.tex}%
\end{tabular}%
}%
\end{table}%

}

\section{Results}

First, we present a comparison between methods (\RQRef{1}) on ad-hoc document retrieval, followed by a per-query comparison between methods (\RQRef{2}) and a combination experiment where we combine latent features with the lexical \QLM{} (\RQRef{3}). We then relate regularities learned by the model to traditional retrieval statistics (\RQRef{4}).

\subsection{Performance of \ModelName{}}
\DoubleColumnTable[Comparison of \ModelName{} with lexical (\QLM{} with \FullDirichlet{} and \FullJelinekMercer{} smoothing) and latent (\DocToVec{}, \WordToVec{}, \LSI{}, \LDA{} and \LSE{}) retrieval models (Section~\ref{sec:retrievalmodels}) on article search benchmarks (Section~\ref{sec:benchmarks}). Significance (Section~\ref{sec:evaluationmeasures}) is computed between \WordToVecSgSi{} and \ModelName{}. Bold values indicate the highest measure value for latent features.]{adhoc}

\RQAnswer{1}{%
Table~\ref{tbl:adhoc_results} shows the retrieval results for ad-hoc document retrieval on the newswire article collections (Section~\ref{sec:benchmarks}).
}{%
We see that \ModelName{} outperforms all other latent rankers on all benchmarks. In particular, \ModelName{} significantly outperforms (\MAPCut{}) the \FullWordToVec{}-based method that weighs word vectors according to self-information (significance is not achieved on \ShortBenchmarkNYT{}). This is an interesting observation as \ModelName{} is trained from scratch without the use of hand-engineered features (i.e., self-information). In the case of \LSE{}, we see that its retrieval effectiveness diminishes. This is due to the large size of the \ShortBenchmarkNYT{} collection (Table~\ref{tbl:adhoc_stats}) and the lack of term specificity within \LSE{} (\S\ref{sec:results:luhn}). Compared to the lexical \QLM{}, \ModelName{} performs better on the \ShortBenchmarkAP{} and \ShortBenchmarkWSJ{} benchmarks. However, it is known that no single ranker performs best on all test sets~\citep{Shaw1994,Liu2011}. In addition, \ModelName{} is a latent model that performs a different type of matching than lexical models. Therefore, we first examine the per-query differences between rankers (\RQRef{2}) and later we will examine the complementary nature of the two types of matching by evaluating combinations of different ranking features (\RQRef{3}).
}

\begin{figure*}[t]
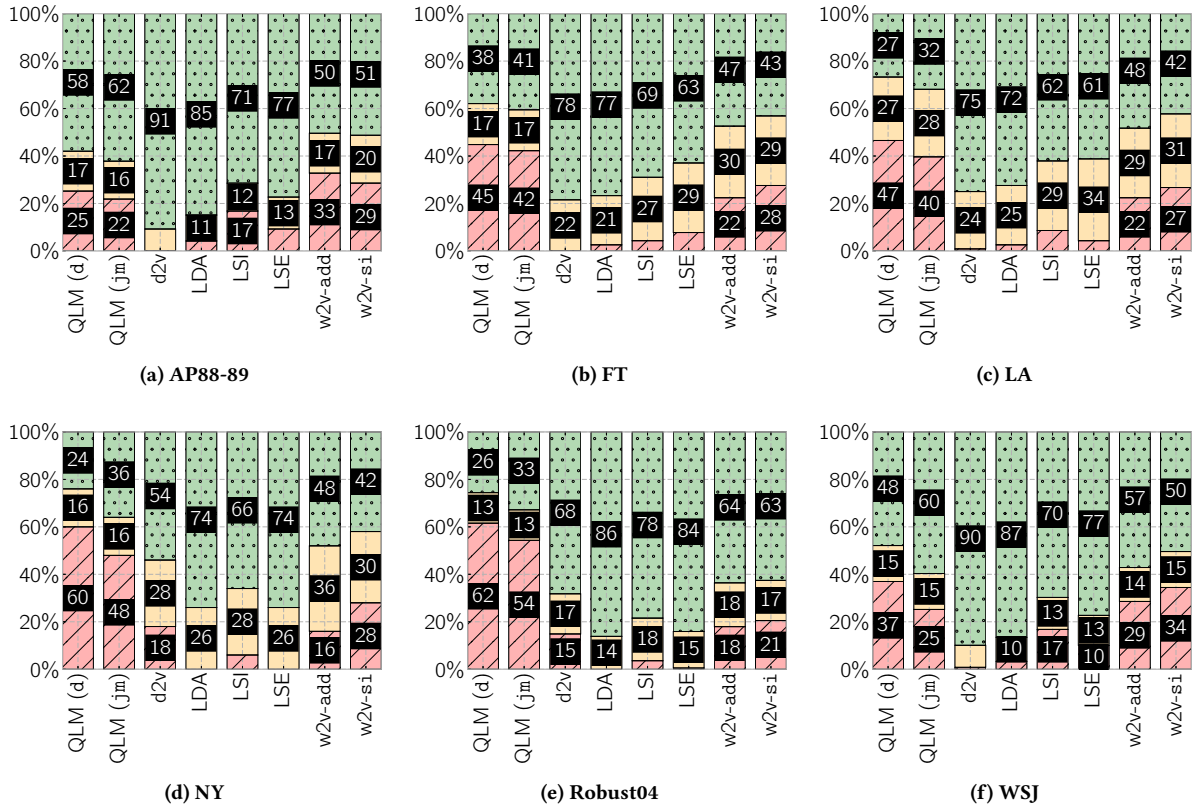


\newcommand{\differences}[2]{%
	\def \inner {%
		\def \PlotPath {resources/analysis/differences/#1/winners.pdf}%
		\IfFileExists{\PlotPath}{%
			\includegraphics[width=\BenchmarkFigureWidth{}]{\PlotPath}}{%
			\resizebox{\BenchmarkFigureWidth{}}{!}{\missingfigure{#1}}}%
	}%
	\ifstrequal{#2}{}{%
	\subfloat{\inner}%
	}{\subfloat[#2\label{fig:differences:#1}]{\inner}}%
}

\centering
\differences{ap_88_89}{\ShortBenchmarkAP{}}%
\hfill%
\differences{ft}{\ShortBenchmarkFT{}}%
\hfill%
\differences{latimes}{\ShortBenchmarkLATimes{}}

\centering
\differences{nyt}{\ShortBenchmarkNYT{}}%
\hfill%
\differences{disk4_disk5_no-cr}{\ShortBenchmarkRobust{}}%
\hfill%
\differences{wsj}{\ShortBenchmarkWSJ{}}%

\smallskip
\caption{Per-query pairwise ranker comparison between \ModelName{} and the \QLMDirichlet{}, \QLMJelinekMercer{}, \DocToVec{}, \WordToVec{}, \LSI{}, \LDA{} and \LSE{} rankers. For every bar, the dotted/green area, solid/orange and red/slashed areas respectively depict the portion of queries for which \ModelName{} outperforms, ties or loses against the other ranker. One ranker outperforms the other if the absolute difference in \MAPCut{} between both rankers exceeds $\delta$.\label{fig:differences}}

\end{figure*}

\DoubleColumnTable[Evaluation of latent features (\DocToVec{}, \WordToVec{}, \LSI{}, \LDA{}, \LSE{} and \ModelName{}) as a complementary signal to \QLMDirichlet{} (Section~\ref{sec:ltr}). Significance (Section~\ref{sec:evaluationmeasures}) is computed between \QLMDirichlet{} + \WordToVecSgSi{} and the best performing combination.]{adhoc_combinations}

\subsection{Query-level analysis}
\label{sec:results:querylevel}
\RQAnswer{2}{%
Figure~\ref{fig:differences} shows the distribution of queries where one individual ranker performs better than the other ($\abs{\Delta \text{\MAPCut{}}} > \delta$).
}{%
We observe similar trends across all benchmarks where \ModelName{} performs best compared to all latent rankers. One competing vector space model, \WordToVecSgSi{}, stands out as it is the strongest baseline and  performs better than \ModelName{} on $20$ to $35\%$ of queries, however, \ModelName{} still beats \WordToVecSgSi{} overall and specifically on $40$ to $55\%$ of queries. Moreover, Figure~\ref{fig:differences} shows us that \QLM{} and \ModelName{} make very different errors. This implies that the combination of \QLM{}, \WordToVecSgSi{} and \ModelName{} might improve performance even further.

While answering \RQRef{1}, we saw that for some benchmarks, latent methods (i.e., \ModelName{}) perform better than lexical methods. While the amount of semantic matching needed depends on various factors, such as the query intent being informational or navigational/transactional \citep{Broder2002taxonomy}, we do see in Figure~\ref{fig:differences} that \ModelName{} performs considerably better amongst latent methods in cases where latent methods perform poorly (e.g., \BenchmarkRobust{} in Figure~\ref{fig:differences:disk4_disk5_no-cr}). Can we shine more light on the difference between \ModelName{} and existing latent methods? We answer this question in the second half of the next section.
}

\newcommand{\DeltaMAPCut}{\Delta{}_{\text{\MAPCut{}}}}
\SingleColumnTable[Correlation coefficients between \TitlestatRelevant{} and $\DeltaMAPCut{}$ between \ModelName{} and the other methods. A positive correlation indicates that \ModelName{} is better at lexical matching, while a negative correlation indicates that \ModelName{} is worse at lexical matching than the alternative. Significance (Section~\ref{sec:evaluationmeasures}) is computed using a permutation test.]{titlestat_correlations}

\subsection{Semantic vs.\ lexical matching}
\RQAnswer{3}{%
Beyond individual rankers, we now also consider combinations of the two best-performing vector space models with \QLM{} \citep{Shaw1994} (see Section~\ref{sec:ltr} for details) in Table~\ref{tbl:adhoc_combinations_results}.
}{%
If we consider the \QLM{} paired with either \WordToVecSgSi{} or \ModelName{}, we see that the combination involving \ModelName{} outperforms the combination with \WordToVecSgSi{} on four out of six benchmarks (\ShortBenchmarkAP{}, \ShortBenchmarkLATimes{}, \ShortBenchmarkNYT{}, \ShortBenchmarkRobust{}). However, Figure~\ref{fig:differences} shows that \ModelName{} and \WordToVecSgSi{} outperform each other on different queries as well. Can we use this difference to our advantage?

The addition of \ModelName{} to the \QLM{} + \WordToVecSgSi{} combination yields an improvement in terms of \MAPCut{} on all benchmarks. Significance is achieved in five out of six benchmarks. In the case of \ShortBenchmarkNYT{} and \BenchmarkRobust{}, the combination of all three rankers (\QLM{} + \WordToVecSgSi{} + \ModelName{}) performs at about the same level as the combination of \QLM{} + \ModelName{}. However, the addition of \ModelName{} to the \QLM{} + \WordToVecSgSi{} combination still creates a significant improvement over just the combination involving \QLM{} and \WordToVec{} only. For \ShortBenchmarkFT{}, the only benchmark where no significance is achieved, we do see that the relative increase in performance nearly doubles from the addition of \ModelName{}. Consequently, we can conclude that the \ModelName{} adds an additional matching signal.

Let us return to the question raised at the end of the previous section (Section~\ref{sec:results:querylevel}): what exactly does the \ModelName{} add in terms of content matching? We investigate this question by determining the amount of semantic matching needed. For each query, we compute \TitlestatRelevant{} (Section~\ref{sec:evaluationmeasures}), the expected normalized overlap between query terms and the terms of relevant document. If \TitlestatRelevant{} is close to \numprint{1.0} for a particular query, then the query requires mostly lexical matching; on the other hand, if \TitlestatRelevant{} is near \numprint{0.0} for a query, then none of the query's relevant document contain the query terms and semantic matching is needed. We continue by examining the per-query pairwise difference ($\DeltaMAPCut{}$) between \ModelName{} and the remaining lexical (\QLM{}) and latent (\DocToVec{}, \LDA{}, \LSI{}, \LSE{}) features. If $\DeltaMAPCut{} > 0$, then \ModelName{} performs better than the other method and vice versa if $\DeltaMAPCut{} < 0$. Table~\ref{tbl:titlestat_correlations_results} shows the Pearson correlation between \TitlestatRelevant{} and $\DeltaMAPCut{}$. A positive correlation, as is the case for \DocToVec{}, \WordToVec{}, \LDA{}, \LSI{} and \LSE{}, indicates that \ModelName{} performs better on queries that require lexical matching. Conversely, a negative correlation, such as observed for both variants of \QLM{}, indicates that \QLM{} performs better on queries that require lexical matching than \ModelName{}. Combining this observation with the conclusion to \RQRef{2} (i.e., \ModelName{} generally improves upon latent methods), we conclude that, in addition to semantic matching, \ModelName{} also performs well in cases where lexical matching is needed and thus contributes a hybrid matching signal.}

\subsection{\ModelName{} and Luhn significance}
\label{sec:results:luhn}

\begin{figure*}[t]
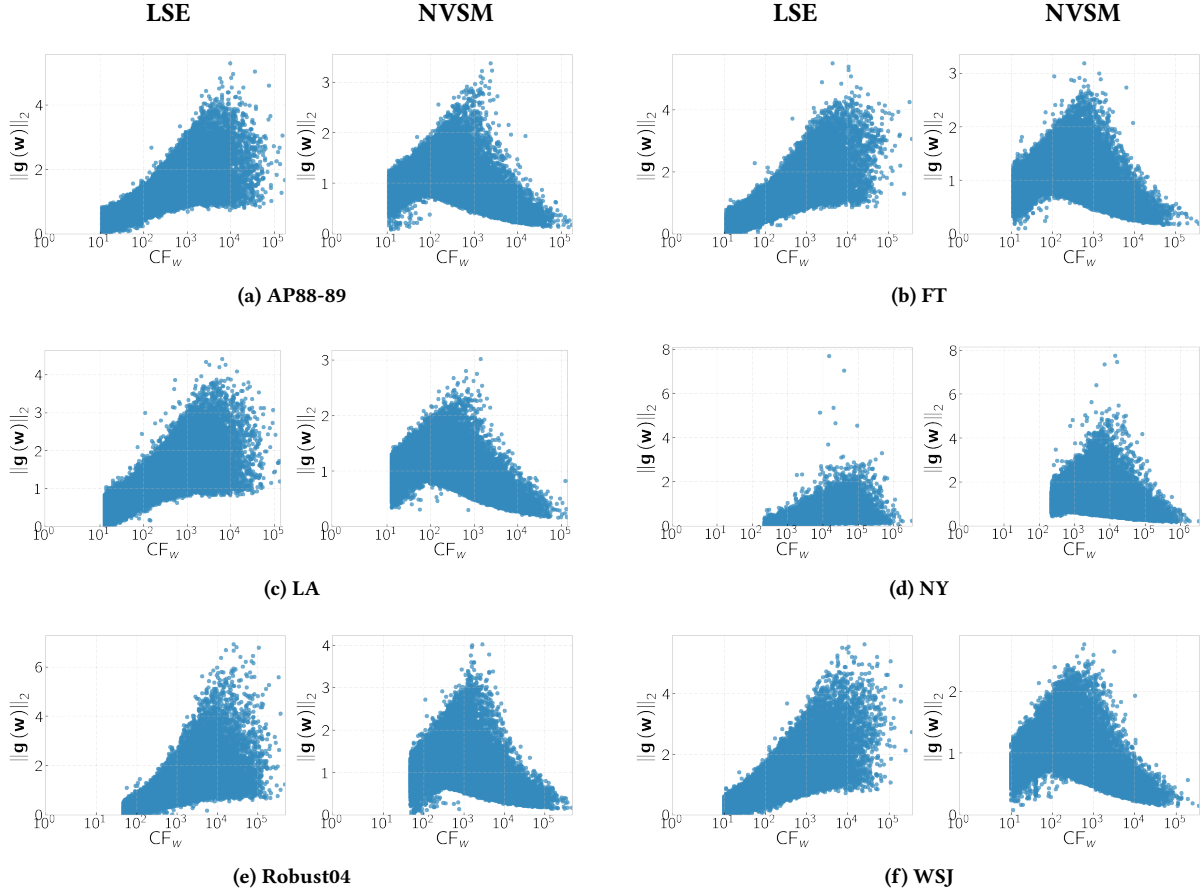


\newcommand{\inner}[1]{}
\newcommand{\regularities}[3]{%
	\renewcommand{\inner}[1]{%
		\def \PlotPath {resources/analysis/model-regularities/#1/##1/#2.png}%
		\IfFileExists{\PlotPath}{%
			\includegraphics[width=\NormFigureWidth{}]{\PlotPath}}{%
			\resizebox{\NormFigureWidth{}}{!}{\missingfigure{#1}}}%
	}%
	\ifstrequal{#3}{}{%
	\subfloat{\inner{lse}   \inner{nvsm}}%
	}{\subfloat[#3\label{fig:regularities:#1}]{\inner{lse}   \inner{nvsm}}}%
}

\newcommand{\NormFigureWidth}{0.235\textwidth}

\centering

\subfloat{\parbox{\NormFigureWidth{}}{\hspace{1.5em}\centering\textbf\LSE{}}   \parbox{\NormFigureWidth{}}{\hspace{1.5em}\centering\textbf\ModelName{}}}%
\hfill%
\subfloat{\parbox{\NormFigureWidth{}}{\hspace{1.5em}\centering\textbf\LSE{}}   \parbox{\NormFigureWidth{}}{\hspace{1.5em}\centering\textbf\ModelName{}}}

\setcounter{subfigure}{0}

\regularities{ap_88_89}{words.collection_frequency_embedding_norm}{\ShortBenchmarkAP{}}%
\hfill%
\regularities{ft}{words.collection_frequency_embedding_norm}{\ShortBenchmarkFT{}}

\regularities{latimes}{words.collection_frequency_embedding_norm}{\ShortBenchmarkLATimes{}}%
\hfill%
\regularities{nyt}{words.collection_frequency_embedding_norm}{\ShortBenchmarkNYT{}}

\regularities{disk4_disk5_no-cr}{words.collection_frequency_embedding_norm}{\ShortBenchmarkRobust{}}%
\hfill%
\regularities{wsj}{words.collection_frequency_embedding_norm}{\ShortBenchmarkWSJ{}}

\bigskip
\caption{Scatter plots of term frequency in the document collections and the L2-norm of the \LSE{} (left) and \ModelName{} (right) representations of these terms. In the case of \LSE{} (left scatter plots for every benchmark), we observe that the L2-norm of term representations grows linearly with the term collection frequency, and consequently, high-frequency terms are of greater importance within \LSE{} representations. For \ModelName{} (right scatter plot for every benchmark), we observe that terms of mid-frequency (middle 50\%) have a statistically significant ($p < 0.01$) higher L2-norm (Section~\ref{sec:evaluationmeasures}) and consequently are of greater importance for retrieval.\label{fig:regularities}}

\end{figure*}

If \ModelName{} performs better at lexical matching than other latent vector space models, does it then also contain regularities associated with term specificity?

\RQAnswer{4}{%
Figure~\ref{fig:regularities} shows the L2-norm of individual term representations for \LSE{} (left scatter plots) and \ModelName{} (right scatter plots).
}{%
\citet{Luhn1958significance} measures the significance of words based on their frequency. They specify a lower and upper frequency cutoff to exclude frequent and infrequent words. For \ModelName{} (scatter plot on the right for every benchmark), we find that infrequent and frequent terms have a statistically significant ($p < 0.01$) smaller L2-norm than terms of medium frequency (Section~\ref{sec:evaluationmeasures}). This observation is further motivated by the shape of the relation between collection frequency in the collection and the L2-norm of term representations in Figure~\ref{fig:regularities}. The key observation that---within \ModelName{} representations---terms of medium frequency are of greater importance (i.e., higher L2-norm) than low- or high-frequency terms closely corresponds to the theory of \citeauthor{Luhn1958significance} significance. Particularly noteworthy is the fact that the \ModelName{} learned this relationship from an unsupervised objective directly, without any notion of relevance. The scatter plots on the left for every benchmark in Figure~\ref{fig:regularities} shows the same analysis for \LSE{} term representations. Unlike with \ModelName{}, we observe that the L2-norm of term representations grows linearly with the term collection frequency, and consequently, high-frequency terms are of greater importance within \LSE{} representations. Therefore, the key difference between \ModelName{} and \LSE{} is that \ModelName{} learns to better encode term specificity.
}
% !TEX root = ./tois2018-adhoc-retrieval.tex

\section{Unsupervised deployment}
\label{sec:analysis:deployment}

In our experiments, we use a validation set for model selection (training iterations and hyperparameters). However, in many cases relevance labels are unavailable. Fortunately, \ModelName{} learns representations of the document collection directly and does not require query-document relevance information. How can we choose values for the hyperparameters of \ModelName{} in the absence of a validation set?
\begin{figure*}[t]
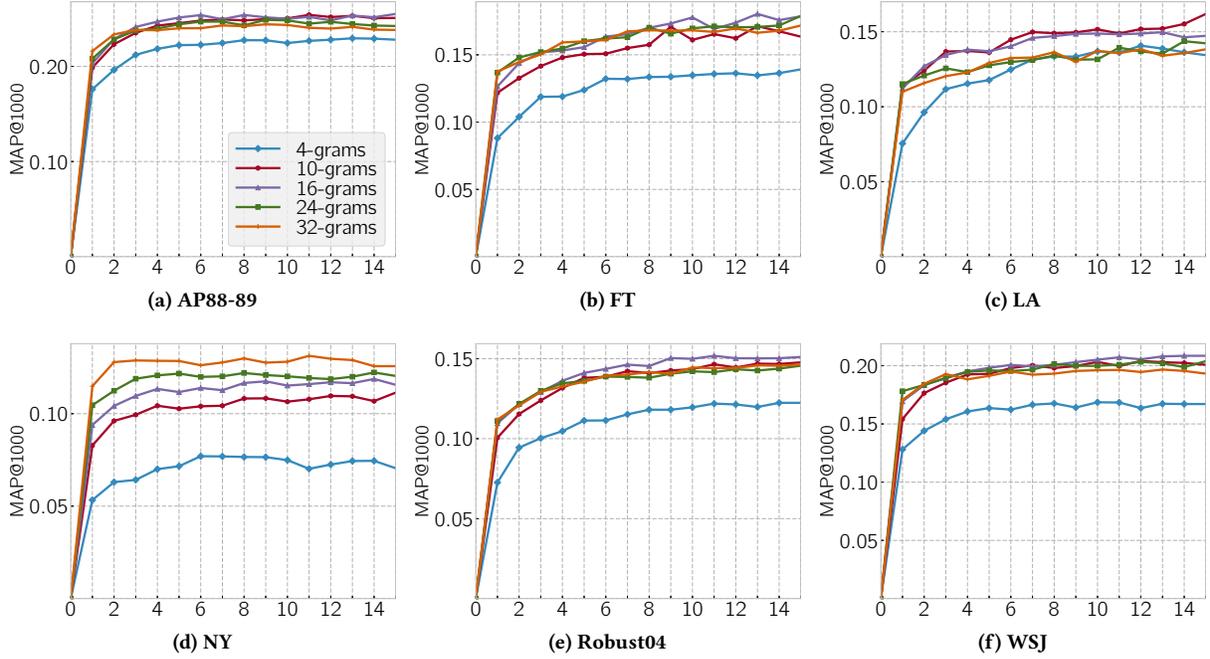


\newcommand{\convergence}[2]{%
    \def \inner {%
        \def \PlotPath {resources/analysis/convergence/#1/#1-qrel_test.pdf}%
        \IfFileExists{\PlotPath}{%
            \includegraphics[width=\BenchmarkFigureWidth{}]{\PlotPath}}{%
            \resizebox{\BenchmarkFigureWidth{}}{!}{\missingfigure{#1}}}%
    }%
    \ifstrequal{#2}{}{%
    \subfloat{\inner}%
    }{\subfloat[#2\label{fig:poolbias:#1}]{\inner}}%
}

\centering
\convergence{ap_88_89}{\ShortBenchmarkAP{}}%
\hfill%
\convergence{ft}{\ShortBenchmarkFT{}}%
\hfill%
\convergence{latimes}{\ShortBenchmarkLATimes{}}

\centering
\convergence{nyt}{\ShortBenchmarkNYT{}}%
\hfill%
\convergence{disk4_disk5_no-cr}{\ShortBenchmarkRobust{}}%
\hfill%
\convergence{wsj}{\ShortBenchmarkWSJ{}}%

\smallskip
\caption{Test set \MAPCut{} as training progresses on article search benchmarks with document space dimensionality $\LatentDocumentDim{} = 256$. We see that \MAPCut{} converges to a fixed performance level with differently-sized $\NGramSize{}$-grams (here we show $\NGramSize{} = 4, 10, 16, 24, 32$; the curves for the remaining values $\NGramSize{} = 6, 8, 12$ are qualitatively similar and omitted to avoid clutter).\label{fig:convergence}}

\vspace*{.25\baselineskip}
\end{figure*}
For the majority of hyperparameters (Section~\ref{sec:retrievalmodels}) we follow the setup of previous work \citep{VanGysel2016products}. We are, however, still tasked with the problem of choosing
\begin{inparaenum}[(a)]
    \item the number of training iterations,
    \item the dimensionality of the document representations $\LatentDocumentDim{}$, and
    \item the size of the $\NGramSize{}$-grams used for training.
\end{inparaenum}
We choose the dimensionality of the document representations $\LatentDocumentDim{} = 256$ as the value was reported to work well for \LSE{} \citep{VanGysel2016products}. Figure~\ref{fig:convergence} shows that \MAPCut{} converges as the number of training iterations increases for different $\NGramSize{}$-gram widths. Therefore, we train \ModelName{} for \numprint{15} iterations and select the last iteration model.
\newcommand{\EnsembleScore}[1]{\Apply{\ScoreFn{}_\text{ensemble}}{\Query{}, \Document{}}}
\newcommand{\Gram}[1]{#1\text{-grams}}
\newcommand{\NGramScoreFn}[1]{\ScoreFn{}_{\Gram{#1}}}
\newcommand{\KGramScoreRandom}[1]{\Apply{\NGramScoreFn{#1}}{\Query{}, \Documents{}}}
\newcommand{\KGramScore}[1]{\Apply{\NGramScoreFn{#1}}{\Query{}, \Document{}}}
\newcommand{\NGramSizes}{\mathcal{N}}
{
    \let\ModelNameSingle\ModelName
    \newcommand{\ModelNameEnsemble}{8 \ModelNameSingle{}s (ensemble)}

    % Temporary redefine.
    \renewcommand{\ModelName}{1 \ModelNameSingle{} (cross-validated)}

    \DoubleColumnTable[Comparison with single cross-validated \ModelNameSingle{} and ensemble of \ModelNameSingle{} through the unsupervised combination of models trained on differently-sized $\NGramSize{}$-grams ($\NGramSizes{} = \{ 2, 4, 8, 10, 12, 16, 24, 32 \}$). Significance (Section~\ref{sec:evaluationmeasures}) is computed between \ModelNameSingle{} and the ensemble of \ModelNameSingle{}.]{adhoc_ensemble}
}
The final remaining question is the choice of $\NGramSize{}$-gram size used during training. This parameter has a big influence on model performance as it determines the amount of context from which semantic relationships are learned. Therefore, we propose to combine different vector spaces trained using different $\NGramSize{}$-gram widths as follows. We write $\NGramSizes{}$ for the set of all $k$ for which we construct an \ModelName{} using $k$-grams. For a given query $\Query{}$, we rank documents $\Document{} \in \Documents{}$ in descending order of:
\begin{equation}
\label{eq:ensemble}
\EnsembleScore{} = \sum_{k \in \NGramSizes{}} \frac{\KGramScore{k} - \mu_{\Gram{k}, \Query{}}}{\sigma_{\Gram{k}, \Query{}}},
\end{equation}
where $\KGramScore{k}$ is Eq.~\ref{eq:score} for \ModelName{} of $\Gram{k}$ and
\begin{equation}
\begin{split}
\mu_{\Gram{k}, \Query{}} & = \ApplySquare{\SampleExpectation}{\KGramScoreRandom{k}} \\
\sigma_{\Gram{k}, \Query{}} & = \sqrt{\ApplySquare{\SampleVariance}{\KGramScoreRandom{k}}},
\end{split}
\end{equation}
denote the sample expectation and sample variance over documents $\Documents{}$ that are estimated on the top-$1000$ documents returned by the individual models, respectively. That is, we rank documents according to the sum of the standardized scores of vector space models trained with different $\NGramSize{}$-gram widths. The score aggregation in Eq.~\ref{eq:ensemble} is performed without any a priori knowledge about the $\NGramSize{}$-gram sizes. Table~\ref{tbl:adhoc_ensemble_results} lists the performance of the unsupervised ensemble, where every model was trained for 15 iterations, against a single cross-validated model. We see that the unsupervised ensemble always outperforms (significantly in terms of \MAPCut{} for all benchmarks except \ShortBenchmarkNYT{}) the singleton model. Hence, we can easily deploy \ModelName{} without any supervision and, surprisingly, it will perform better than individual models optimized on a validation set.

% !TEX root = ./tois2018-adhoc-retrieval.tex

\section{Conclusions}

We have proposed the \FullModelName{} (\ModelName{}) that learns representations of a document collection in an unsupervised manner. 
We have shown that \ModelName{} performs better than existing latent vector space/bag-of-words approaches. \ModelName{} performs lexical and semantic matching in a latent space. \ModelName{} provides a complementary signal to lexical language models. In addition, we have shown that \ModelName{} automatically learns a notion of term specificity. Finally, we have given advice on how to select values for the hyperparameters of \ModelName{}. Interestingly, an unsupervised ensemble of multiple models trained with different hyperparameters performs better than a single cross-validated model.

The evidence that \ModelName{} provides a notion of lexical matching tells us that latent vector space models are not limited to only semantic matching. While the framework presented in this paper focuses on a single unsupervised objective, additional objectives (i.e., document/document or query/document similarity) can be incorporated to improve retrieval performance.

\LSE{} \citep{VanGysel2016products} improved the learning time complexity of earlier entity retrieval models \citep{VanGysel2016experts} such that they scale to ${\sim}100\text{k}$ retrievable items (i.e., entities). However, as shown in Table~\ref{tbl:adhoc_results}, \LSE{} performs poorly on article retrieval benchmarks. In this paper, we extend \LSE{} and learn vector spaces of ${\sim}500\text{k}$ documents that perform better than existing latent vector spaces. As mentioned in the introduction, the main challenge for latent vector spaces is their limited scalability to large document collections due to space complexity. The observation that retrieval is not only impacted by the vector space representation of the relevant document, but also of the documents surrounding it, raises non-trivial questions regarding the distribution of document vectors over multiple machines. While there have been efforts towards distributed training of neural models, the application of distributed learning algorithms is left for future work. The unsupervised objective that learns from word sequences is limited by its inability to deal with very short documents. While this makes the unsupervised objective less applicable in domains such as web search, unsupervised bag-of-words approaches have the opposite problem of degrading performance when used to search over long documents. With respect to incremental indexing, there is currently no theoretically sound way to obtain representations for new documents that were added to the collection after the initial estimation of a \ModelName{}. In the case of \LDA{} or \LSI{}, representations for new documents can be obtained by transforming bag-of-words vectors to the latent space. However, as the \LDA{}/\LSI{} transformation to the latent space is not updated after estimating the \LDA{}/\LSI{} model using the initial set of documents, this procedure can be catastrophic when topic drift occurs. For \FullDocToVec{}, one way to obtain a representation for a previously-unseen document is to keep all parameters fixed and train the representation of the new document using the standard training algorithm \citep{Rehurek2010gensim}. This approach can also be used in the case of \LSE{} or \ModelName{}. However, there are no guarantees that the obtained representation will be of desirable quality. In addition, the same problem remains as with the bag-of-words methods. That is, the previously-mentioned incremental updating mechanism is likely to fail when topic drift occurs.

We hope that our work and the insights resulting from it inspires others to further develop unsupervised neural retrieval models. Future work includes adding additional model expressiveness through depth or width, in-depth analysis of the various components on NVSM, analysis of the learned representations and their combination in various combination frameworks (\S\ref{sec:related:unsupervised}) and engineering challenges regarding the scalability of the training procedure.

\begin{acks}
We thank Adith Swaminathan, Alexey Borisov, Tom Kenter, Hosein Azarbonyad, Mostafa Dehghani, Nikos Voskarides and Adam Holmes and the anonymous reviewers for their helpful comments.
\end{acks}

\bibliographystyle{ACM-Reference-Format}
\bibliography{tois2018-adhoc-retrieval}


\begin{thebibliography}{87}

%%% the Web address.

\ifx \showCODEN    \undefined \def \showCODEN     #1{\unskip}     \fi
\ifx \showDOI      \undefined \def \showDOI       #1{#1}\fi
\ifx \showISBNx    \undefined \def \showISBNx     #1{\unskip}     \fi
\ifx \showISBNxiii \undefined \def \showISBNxiii  #1{\unskip}     \fi
\ifx \showISSN     \undefined \def \showISSN      #1{\unskip}     \fi
\ifx \showLCCN     \undefined \def \showLCCN      #1{\unskip}     \fi
\ifx \shownote     \undefined \def \shownote      #1{#1}          \fi
\ifx \showarticletitle \undefined \def \showarticletitle #1{#1}   \fi
\ifx \showURL      \undefined \def \showURL       {\relax}        \fi
% invisible to TeX
\providecommand\bibfield[2]{#2}
\providecommand\bibinfo[2]{#2}
\providecommand\natexlab[1]{#1}
\providecommand\showeprint[2][]{arXiv:#2}

\bibitem[\protect\citeauthoryear{Abadi, Agarwal, Barham, Brevdo, Chen, Citro,
  Corrado, Davis, Dean, Devin, Ghemawat, Goodfellow, Harp, Irving, Isard, Jia,
  Jozefowicz, Kaiser, Kudlur, Levenberg, Man\'{e}, Monga, Moore, Murray,
  Chris~Olah, Schuster, Shlens, Steiner, Sutskever, Talwar, Tucker, Vanhoucke,
  Vasudevan, Vi\'{e}gas, Vinyals, Warden, Wattenberg, Wicke, Yu, and
  Xiaoqiang}{Abadi et~al\mbox{.}}{2015}]%
        {Tensorflow2015whitepaper}
\bibfield{author}{\bibinfo{person}{Mart\'{\i}n Abadi}, \bibinfo{person}{Ashish
  Agarwal}, \bibinfo{person}{Paul Barham}, \bibinfo{person}{Eugene Brevdo},
  \bibinfo{person}{Zhifeng Chen}, \bibinfo{person}{Craig Citro},
  \bibinfo{person}{Greg~S. Corrado}, \bibinfo{person}{Andy Davis},
  \bibinfo{person}{Jeffrey Dean}, \bibinfo{person}{Matthieu Devin},
  \bibinfo{person}{Sanjay Ghemawat}, \bibinfo{person}{Ian Goodfellow},
  \bibinfo{person}{Andrew Harp}, \bibinfo{person}{Geoffrey Irving},
  \bibinfo{person}{Michael Isard}, \bibinfo{person}{Yangqing Jia},
  \bibinfo{person}{Rafal Jozefowicz}, \bibinfo{person}{Lukasz Kaiser},
  \bibinfo{person}{Manjunath Kudlur}, \bibinfo{person}{Josh Levenberg},
  \bibinfo{person}{Dan Man\'{e}}, \bibinfo{person}{Rajat Monga},
  \bibinfo{person}{Sherry Moore}, \bibinfo{person}{Derek Murray},
  \bibinfo{person}{Derek Chris~Olah}, \bibinfo{person}{Mike Schuster},
  \bibinfo{person}{Jonathon Shlens}, \bibinfo{person}{Benoi Steiner},
  \bibinfo{person}{Ilya Sutskever}, \bibinfo{person}{Kunal Talwar},
  \bibinfo{person}{Paul Tucker}, \bibinfo{person}{Vincent Vanhoucke},
  \bibinfo{person}{Vijay Vasudevan}, \bibinfo{person}{Fernanda Vi\'{e}gas},
  \bibinfo{person}{Oriol Vinyals}, \bibinfo{person}{Pete Warden},
  \bibinfo{person}{Martin Wattenberg}, \bibinfo{person}{Martin Wicke},
  \bibinfo{person}{Yuan Yu}, {and} \bibinfo{person}{Zheng Xiaoqiang}.}
  \bibinfo{year}{2015}\natexlab{}.
\newblock \bibinfo{title}{{TensorFlow}: Large-scale machine learning on
  heterogeneous systems}.
\newblock   (\bibinfo{year}{2015}).
\newblock


\bibitem[\protect\citeauthoryear{Ai, Yang, Guo, and Croft}{Ai
  et~al\mbox{.}}{2016a}]%
        {Ai2016doc2vecanalysis}
\bibfield{author}{\bibinfo{person}{Qingyao Ai}, \bibinfo{person}{Liu Yang},
  \bibinfo{person}{Jiafeng Guo}, {and} \bibinfo{person}{W.~Bruce Croft}.}
  \bibinfo{year}{2016}\natexlab{a}.
\newblock \showarticletitle{Analysis of the paragraph vector model for
  information retrieval}. In \bibinfo{booktitle}{\emph{ICTIR}}. ACM,
  \bibinfo{pages}{133--142}.
\newblock


\bibitem[\protect\citeauthoryear{Ai, Yang, Guo, and Croft}{Ai
  et~al\mbox{.}}{2016b}]%
        {Ai2016doc2veclm}
\bibfield{author}{\bibinfo{person}{Qingyao Ai}, \bibinfo{person}{Liu Yang},
  \bibinfo{person}{Jiafeng Guo}, {and} \bibinfo{person}{W~Bruce Croft}.}
  \bibinfo{year}{2016}\natexlab{b}.
\newblock \showarticletitle{Improving language estimation with the paragraph
  vector model for ad-hoc retrieval}. In \bibinfo{booktitle}{\emph{SIGIR}}.
  ACM, \bibinfo{pages}{869--872}.
\newblock


\bibitem[\protect\citeauthoryear{Allan, Harman, Kanoulas, Li, Van~Gysel, and
  Voorhees}{Allan et~al\mbox{.}}{2017}]%
        {Allan2017treccommon}
\bibfield{author}{\bibinfo{person}{James Allan}, \bibinfo{person}{Donna
  Harman}, \bibinfo{person}{Evangelos Kanoulas}, \bibinfo{person}{Dan Li},
  \bibinfo{person}{Christophe Van~Gysel}, {and} \bibinfo{person}{Ellen
  Voorhees}.} \bibinfo{year}{2017}\natexlab{}.
\newblock \showarticletitle{TREC 2017 Common Core Track Overview}. In
  \bibinfo{booktitle}{\emph{TREC}}.
\newblock


\bibitem[\protect\citeauthoryear{Baroni, Dinu, and Kruszewski}{Baroni
  et~al\mbox{.}}{2014}]%
        {Baroni2014}
\bibfield{author}{\bibinfo{person}{Marco Baroni}, \bibinfo{person}{Georgiana
  Dinu}, {and} \bibinfo{person}{Germ{\'a}n Kruszewski}.}
  \bibinfo{year}{2014}\natexlab{}.
\newblock \showarticletitle{{Don't count, predict! A systematic comparison of
  context-counting vs. context-predicting semantic vectors}}. In
  \bibinfo{booktitle}{\emph{ACL}}. \bibinfo{pages}{238--247}.
\newblock


\bibitem[\protect\citeauthoryear{Bengio, Ducharme, Vincent, and Janvin}{Bengio
  et~al\mbox{.}}{2003}]%
        {Bengio2003}
\bibfield{author}{\bibinfo{person}{Yoshua Bengio}, \bibinfo{person}{R\'{e}jean
  Ducharme}, \bibinfo{person}{Pascal Vincent}, {and} \bibinfo{person}{Christian
  Janvin}.} \bibinfo{year}{2003}\natexlab{}.
\newblock \showarticletitle{A neural probabilistic language model}.
\newblock \bibinfo{journal}{\emph{JMLR}}  \bibinfo{volume}{3}
  (\bibinfo{year}{2003}), \bibinfo{pages}{1137--1155}.
\newblock


\bibitem[\protect\citeauthoryear{Blei, Ng, and Jordan}{Blei
  et~al\mbox{.}}{2003}]%
        {Blei2003}
\bibfield{author}{\bibinfo{person}{David~M Blei}, \bibinfo{person}{Andrew~Y
  Ng}, {and} \bibinfo{person}{Michael~I Jordan}.}
  \bibinfo{year}{2003}\natexlab{}.
\newblock \showarticletitle{Latent dirichlet allocation}.
\newblock \bibinfo{journal}{\emph{JMLR}}  \bibinfo{volume}{3}
  (\bibinfo{year}{2003}), \bibinfo{pages}{993--1022}.
\newblock


\bibitem[\protect\citeauthoryear{Borisov, Markov, de~Rijke, and
  Serdyukov}{Borisov et~al\mbox{.}}{2016a}]%
        {Borisov2016contextaware}
\bibfield{author}{\bibinfo{person}{Alexey Borisov}, \bibinfo{person}{Ilya
  Markov}, \bibinfo{person}{Maarten de Rijke}, {and} \bibinfo{person}{Pavel
  Serdyukov}.} \bibinfo{year}{2016}\natexlab{a}.
\newblock \showarticletitle{A context-aware time model for web search}. In
  \bibinfo{booktitle}{\emph{SIGIR}}. \bibinfo{publisher}{ACM},
  \bibinfo{pages}{205--214}.
\newblock


\bibitem[\protect\citeauthoryear{Borisov, Markov, de~Rijke, and
  Serdyukov}{Borisov et~al\mbox{.}}{2016b}]%
        {Borisov2016click}
\bibfield{author}{\bibinfo{person}{Alexey Borisov}, \bibinfo{person}{Ilya
  Markov}, \bibinfo{person}{Maarten de Rijke}, {and} \bibinfo{person}{Pavel
  Serdyukov}.} \bibinfo{year}{2016}\natexlab{b}.
\newblock \showarticletitle{A neural click model for web search}. In
  \bibinfo{booktitle}{\emph{WWW}}. International World Wide Web Conferences
  Steering Committee, \bibinfo{pages}{531--541}.
\newblock


\bibitem[\protect\citeauthoryear{Boytsov, Novak, Malkov, and Eric}{Boytsov
  et~al\mbox{.}}{2016}]%
        {Boytsov2016knn}
\bibfield{author}{\bibinfo{person}{Leonid Boytsov}, \bibinfo{person}{David
  Novak}, \bibinfo{person}{Yury Malkov}, {and} \bibinfo{person}{Nyberg Eric}.}
  \bibinfo{year}{2016}\natexlab{}.
\newblock \showarticletitle{Off the beaten path: Let's replace term-based
  retrieval with k-NN search}. In \bibinfo{booktitle}{\emph{CIKM}}.
  \bibinfo{pages}{1099--1108}.
\newblock


\bibitem[\protect\citeauthoryear{Broder}{Broder}{2002}]%
        {Broder2002taxonomy}
\bibfield{author}{\bibinfo{person}{Andrei Broder}.}
  \bibinfo{year}{2002}\natexlab{}.
\newblock \showarticletitle{A taxonomy of web search}.
\newblock \bibinfo{journal}{\emph{SIGIR forum}} \bibinfo{volume}{36},
  \bibinfo{number}{2} (\bibinfo{year}{2002}), \bibinfo{pages}{3--10}.
\newblock


\bibitem[\protect\citeauthoryear{Buckley, Dimmick, Soboroff, and
  Voorhees}{Buckley et~al\mbox{.}}{2007}]%
        {Buckley2007bias}
\bibfield{author}{\bibinfo{person}{Chris Buckley}, \bibinfo{person}{Darrin
  Dimmick}, \bibinfo{person}{Ian Soboroff}, {and} \bibinfo{person}{Ellen
  Voorhees}.} \bibinfo{year}{2007}\natexlab{}.
\newblock \showarticletitle{Bias and the limits of pooling for large
  collections}.
\newblock \bibinfo{journal}{\emph{Information retrieval}} \bibinfo{volume}{10},
  \bibinfo{number}{6} (\bibinfo{year}{2007}), \bibinfo{pages}{491--508}.
\newblock


\bibitem[\protect\citeauthoryear{Burges, Shaked, Renshaw, Lazier, Deeds,
  Hamilton, and Hullender}{Burges et~al\mbox{.}}{2005}]%
        {Burges2005ranknet}
\bibfield{author}{\bibinfo{person}{Chris Burges}, \bibinfo{person}{Tal Shaked},
  \bibinfo{person}{Erin Renshaw}, \bibinfo{person}{Ari Lazier},
  \bibinfo{person}{Matt Deeds}, \bibinfo{person}{Nicole Hamilton}, {and}
  \bibinfo{person}{Greg Hullender}.} \bibinfo{year}{2005}\natexlab{}.
\newblock \showarticletitle{Learning to rank using gradient descent}. In
  \bibinfo{booktitle}{\emph{ICML}}. ACM, \bibinfo{pages}{89--96}.
\newblock


\bibitem[\protect\citeauthoryear{Chapelle and Zhang}{Chapelle and
  Zhang}{2009}]%
        {Chapelle2009dbn}
\bibfield{author}{\bibinfo{person}{Olivier Chapelle} {and} \bibinfo{person}{Ya
  Zhang}.} \bibinfo{year}{2009}\natexlab{}.
\newblock \showarticletitle{A dynamic bayesian network click model for web
  search ranking}. In \bibinfo{booktitle}{\emph{WWW}}. ACM,
  \bibinfo{pages}{1--10}.
\newblock


\bibitem[\protect\citeauthoryear{Chen}{Chen}{2017}]%
        {Chen2017corruption}
\bibfield{author}{\bibinfo{person}{Minmin Chen}.}
  \bibinfo{year}{2017}\natexlab{}.
\newblock \showarticletitle{Efficient vector representation for documents
  through corruption}. In \bibinfo{booktitle}{\emph{ICLR}}.
\newblock


\bibitem[\protect\citeauthoryear{Collobert, Weston, Bottou, Karlen,
  Kavukcuoglu, and Kuksa}{Collobert et~al\mbox{.}}{2011}]%
        {Collobert2011scratch}
\bibfield{author}{\bibinfo{person}{Ronan Collobert}, \bibinfo{person}{Jason
  Weston}, \bibinfo{person}{Leon Bottou}, \bibinfo{person}{Michael Karlen},
  \bibinfo{person}{Koray Kavukcuoglu}, {and} \bibinfo{person}{Pavel Kuksa}.}
  \bibinfo{year}{2011}\natexlab{}.
\newblock \showarticletitle{Natural language processing (almost) from scratch}.
\newblock \bibinfo{journal}{\emph{JMLR}} \bibinfo{volume}{12},
  \bibinfo{number}{Aug} (\bibinfo{year}{2011}), \bibinfo{pages}{2493--2537}.
\newblock
\showISSN{1532-4435}


\bibitem[\protect\citeauthoryear{Cover and Thomas}{Cover and Thomas}{2012}]%
        {Cover2012informationtheory}
\bibfield{author}{\bibinfo{person}{Thomas~M Cover} {and} \bibinfo{person}{Joy~A
  Thomas}.} \bibinfo{year}{2012}\natexlab{}.
\newblock \bibinfo{booktitle}{\emph{Elements of information theory}}.
\newblock \bibinfo{publisher}{John Wiley \& Sons}.
\newblock


\bibitem[\protect\citeauthoryear{Craswell, Croft, Guo, Mitra, and
  de~Rijke}{Craswell et~al\mbox{.}}{2016}]%
        {Craswell2016neuir}
\bibfield{author}{\bibinfo{person}{Nick Craswell}, \bibinfo{person}{W.~Bruce
  Croft}, \bibinfo{person}{Jiafeng Guo}, \bibinfo{person}{Bhaskar Mitra}, {and}
  \bibinfo{person}{Maarten de Rijke}.} \bibinfo{year}{2016}\natexlab{}.
\newblock \showarticletitle{Neu-IR: The SIGIR 2016 workshop on neural
  information retrieval}. In \bibinfo{booktitle}{\emph{SIGIR}}.
  \bibinfo{publisher}{ACM}, \bibinfo{pages}{1245--1246}.
\newblock


\bibitem[\protect\citeauthoryear{Deerwester, Dumais, Furnas, Landauer, and
  Harshman}{Deerwester et~al\mbox{.}}{1990}]%
        {Deerwester1990}
\bibfield{author}{\bibinfo{person}{Scott~C. Deerwester},
  \bibinfo{person}{Susan~T Dumais}, \bibinfo{person}{George~W. Furnas},
  \bibinfo{person}{Thomas~K. Landauer}, {and} \bibinfo{person}{Richard~A.
  Harshman}.} \bibinfo{year}{1990}\natexlab{}.
\newblock \showarticletitle{Indexing by latent semantic analysis}.
\newblock \bibinfo{journal}{\emph{JASIS}} \bibinfo{volume}{41},
  \bibinfo{number}{6} (\bibinfo{year}{1990}), \bibinfo{pages}{391--407}.
\newblock


\bibitem[\protect\citeauthoryear{Deng, He, and Gao}{Deng et~al\mbox{.}}{2013}]%
        {Deng2013}
\bibfield{author}{\bibinfo{person}{Li Deng}, \bibinfo{person}{Xiaodong He},
  {and} \bibinfo{person}{Jianfeng Gao}.} \bibinfo{year}{2013}\natexlab{}.
\newblock \showarticletitle{{Deep stacking networks for information
  retrieval}}. In \bibinfo{booktitle}{\emph{ICASSP}}.
  \bibinfo{pages}{3153--3157}.
\newblock


\bibitem[\protect\citeauthoryear{Dumais}{Dumais}{1995}]%
        {Dumais95lsi}
\bibfield{author}{\bibinfo{person}{Susan~T. Dumais}.}
  \bibinfo{year}{1995}\natexlab{}.
\newblock \showarticletitle{Latent semantic indexing (LSI): TREC-3 Report}. In
  \bibinfo{booktitle}{\emph{TREC}}. \bibinfo{publisher}{NIST},
  \bibinfo{pages}{219--230}.
\newblock


\bibitem[\protect\citeauthoryear{Folk, Heber, Koziol, Pourmal, and
  Robinson}{Folk et~al\mbox{.}}{2011}]%
        {Folk2011hdf5}
\bibfield{author}{\bibinfo{person}{Mike Folk}, \bibinfo{person}{Gerd Heber},
  \bibinfo{person}{Quincey Koziol}, \bibinfo{person}{Elena Pourmal}, {and}
  \bibinfo{person}{Dana Robinson}.} \bibinfo{year}{2011}\natexlab{}.
\newblock \showarticletitle{An overview of the HDF5 technology suite and its
  applications}. In \bibinfo{booktitle}{\emph{EDBT/ICDT Workshop on Array
  Databases}}. ACM, \bibinfo{pages}{36--47}.
\newblock


\bibitem[\protect\citeauthoryear{Ganguly, Roy, Mitra, and Jones}{Ganguly
  et~al\mbox{.}}{2015}]%
        {Ganguly2015generalizedlm}
\bibfield{author}{\bibinfo{person}{Debasis Ganguly}, \bibinfo{person}{Dwaipayan
  Roy}, \bibinfo{person}{Mandar Mitra}, {and} \bibinfo{person}{Gareth~JF
  Jones}.} \bibinfo{year}{2015}\natexlab{}.
\newblock \showarticletitle{Word embedding based generalized language model for
  information retrieval}. In \bibinfo{booktitle}{\emph{SIGIR}}. ACM,
  \bibinfo{pages}{795--798}.
\newblock


\bibitem[\protect\citeauthoryear{Garcia, Debreuve, and Barlaud}{Garcia
  et~al\mbox{.}}{2008}]%
        {Garcia2008fastknn}
\bibfield{author}{\bibinfo{person}{Vincent Garcia}, \bibinfo{person}{Eric
  Debreuve}, {and} \bibinfo{person}{Michel Barlaud}.}
  \bibinfo{year}{2008}\natexlab{}.
\newblock \showarticletitle{Fast k nearest neighbor search using GPU}. In
  \bibinfo{booktitle}{\emph{CVPRW}}. IEEE, \bibinfo{pages}{1--6}.
\newblock


\bibitem[\protect\citeauthoryear{Graves and Jaitly}{Graves and Jaitly}{2014}]%
        {Graves2014speech}
\bibfield{author}{\bibinfo{person}{Alex Graves} {and} \bibinfo{person}{Navdeep
  Jaitly}.} \bibinfo{year}{2014}\natexlab{}.
\newblock \showarticletitle{Towards end-to-end speech recognition with
  recurrent neural networks}. In \bibinfo{booktitle}{\emph{ICML}}.
  \bibinfo{pages}{1764--1772}.
\newblock


\bibitem[\protect\citeauthoryear{Gulcehre, Moczulski, Denil, and
  Bengio}{Gulcehre et~al\mbox{.}}{2016}]%
        {Gulcehre2016noisy}
\bibfield{author}{\bibinfo{person}{Caglar Gulcehre}, \bibinfo{person}{Marcin
  Moczulski}, \bibinfo{person}{Misha Denil}, {and} \bibinfo{person}{Yoshua
  Bengio}.} \bibinfo{year}{2016}\natexlab{}.
\newblock \showarticletitle{Noisy activation functions}.
\newblock \bibinfo{journal}{\emph{arXiv preprint arXiv:1603.00391}}
  (\bibinfo{year}{2016}).
\newblock


\bibitem[\protect\citeauthoryear{Guo, Fan, Ai, and Croft}{Guo
  et~al\mbox{.}}{2016a}]%
        {Guo2016relevance}
\bibfield{author}{\bibinfo{person}{Jiafeng Guo}, \bibinfo{person}{Yixing Fan},
  \bibinfo{person}{Qingyao Ai}, {and} \bibinfo{person}{W~Bruce Croft}.}
  \bibinfo{year}{2016}\natexlab{a}.
\newblock \showarticletitle{A deep relevance matching model for ad-hoc
  retrieval}. In \bibinfo{booktitle}{\emph{CIKM}}. ACM,
  \bibinfo{pages}{55--64}.
\newblock


\bibitem[\protect\citeauthoryear{Guo, Fan, Ai, and Croft}{Guo
  et~al\mbox{.}}{2016b}]%
        {Guo2016wordtransport}
\bibfield{author}{\bibinfo{person}{Jiafeng Guo}, \bibinfo{person}{Yixing Fan},
  \bibinfo{person}{Qingyao Ai}, {and} \bibinfo{person}{W.~Bruce Croft}.}
  \bibinfo{year}{2016}\natexlab{b}.
\newblock \showarticletitle{Semantic matching by non-linear word transportation
  for information retrieval}. In \bibinfo{booktitle}{\emph{CIKM}}.
  \bibinfo{publisher}{ACM}, \bibinfo{pages}{701--710}.
\newblock


\bibitem[\protect\citeauthoryear{Gutmann and Hyv{\"a}rinen}{Gutmann and
  Hyv{\"a}rinen}{2010}]%
        {Gutmann2010}
\bibfield{author}{\bibinfo{person}{Michael Gutmann} {and} \bibinfo{person}{Aapo
  Hyv{\"a}rinen}.} \bibinfo{year}{2010}\natexlab{}.
\newblock \showarticletitle{Noise-contrastive estimation: A new estimation
  principle for unnormalized statistical models}. In
  \bibinfo{booktitle}{\emph{AISTATS}}. \bibinfo{pages}{297--304}.
\newblock


\bibitem[\protect\citeauthoryear{Harman}{Harman}{1992}]%
        {Harman1992tipster}
\bibfield{author}{\bibinfo{person}{Donna Harman}.}
  \bibinfo{year}{1992}\natexlab{}.
\newblock \showarticletitle{The DARPA TIPSTER Project}.
\newblock \bibinfo{journal}{\emph{SIGIR Forum}} \bibinfo{volume}{26},
  \bibinfo{number}{2} (\bibinfo{date}{Oct.} \bibinfo{year}{1992}),
  \bibinfo{pages}{26--28}.
\newblock


\bibitem[\protect\citeauthoryear{Harman}{Harman}{1993}]%
        {Harman1993document}
\bibfield{author}{\bibinfo{person}{Donna Harman}.}
  \bibinfo{year}{1993}\natexlab{}.
\newblock \showarticletitle{Document detection data preparation}. In
  \bibinfo{booktitle}{\emph{TIPSTER TEXT PROGRAM: PHASE I: Proceedings of a
  Workshop held at Fredricksburg, Virginia, September 19-23, 1993}}. ACL,
  \bibinfo{pages}{17--31}.
\newblock


\bibitem[\protect\citeauthoryear{Harman and Voorhees}{Harman and
  Voorhees}{1996}]%
        {Harman1996trec5}
\bibfield{author}{\bibinfo{person}{Donna Harman} {and} \bibinfo{person}{Ellen
  Voorhees}.} \bibinfo{year}{1996}\natexlab{}.
\newblock \showarticletitle{Overview of the fifth text retrieval conference}.
  In \bibinfo{booktitle}{\emph{TREC-5}}. \bibinfo{pages}{500--238}.
\newblock


\bibitem[\protect\citeauthoryear{Hofmann}{Hofmann}{1999}]%
        {Hofmann1999}
\bibfield{author}{\bibinfo{person}{Thomas Hofmann}.}
  \bibinfo{year}{1999}\natexlab{}.
\newblock \showarticletitle{Probabilistic latent semantic indexing}. In
  \bibinfo{booktitle}{\emph{SIGIR}}. ACM, \bibinfo{pages}{50--57}.
\newblock


\bibitem[\protect\citeauthoryear{Huang, Urbana, He, Gao, Deng, Acero, and
  Heck}{Huang et~al\mbox{.}}{2013}]%
        {Huang2013}
\bibfield{author}{\bibinfo{person}{Po-sen Huang},
  \bibinfo{person}{N~Mathews~Ave Urbana}, \bibinfo{person}{Xiaodong He},
  \bibinfo{person}{Jianfeng Gao}, \bibinfo{person}{Li Deng},
  \bibinfo{person}{Alex Acero}, {and} \bibinfo{person}{Larry Heck}.}
  \bibinfo{year}{2013}\natexlab{}.
\newblock \showarticletitle{Learning deep structured semantic models for web
  search using clickthrough data}. In \bibinfo{booktitle}{\emph{CIKM}}.
  \bibinfo{pages}{2333--2338}.
\newblock


\bibitem[\protect\citeauthoryear{Ioffe and Szegedy}{Ioffe and Szegedy}{2015}]%
        {Ioffe15bn}
\bibfield{author}{\bibinfo{person}{Sergey Ioffe} {and}
  \bibinfo{person}{Christian Szegedy}.} \bibinfo{year}{2015}\natexlab{}.
\newblock \showarticletitle{Batch normalization: Accelerating deep network
  training by reducing internal covariate shift}.
\newblock \bibinfo{journal}{\emph{CoRR}}  \bibinfo{volume}{abs/1502.03167}
  (\bibinfo{year}{2015}).
\newblock
\urldef\tempurl%
\url{http://arxiv.org/abs/1502.03167}
\showURL{%
\tempurl}


\bibitem[\protect\citeauthoryear{Joachims}{Joachims}{2002}]%
        {Joachims2002svm}
\bibfield{author}{\bibinfo{person}{Thorsten Joachims}.}
  \bibinfo{year}{2002}\natexlab{}.
\newblock \showarticletitle{Optimizing search engines using clickthrough data}.
  In \bibinfo{booktitle}{\emph{SIGKDD}}. ACM, \bibinfo{pages}{133--142}.
\newblock


\bibitem[\protect\citeauthoryear{Jozefowicz, Zaremba, and Sutskever}{Jozefowicz
  et~al\mbox{.}}{2015}]%
        {Jozefowicz2015rnn}
\bibfield{author}{\bibinfo{person}{Rafal Jozefowicz}, \bibinfo{person}{Wojciech
  Zaremba}, {and} \bibinfo{person}{Ilya Sutskever}.}
  \bibinfo{year}{2015}\natexlab{}.
\newblock \showarticletitle{An empirical exploration of recurrent network
  architectures}. In \bibinfo{booktitle}{\emph{ICML}}.
  \bibinfo{pages}{2342--2350}.
\newblock


\bibitem[\protect\citeauthoryear{Kenter, Borisov, and de~Rijke}{Kenter
  et~al\mbox{.}}{2016}]%
        {Kenter2016siamese}
\bibfield{author}{\bibinfo{person}{Tom Kenter}, \bibinfo{person}{Alexey
  Borisov}, {and} \bibinfo{person}{Maarten de Rijke}.}
  \bibinfo{year}{2016}\natexlab{}.
\newblock \showarticletitle{Siamese CBOW: Optimizing word embeddings for
  sentence representations}. In \bibinfo{booktitle}{\emph{ACL}}.
  \bibinfo{pages}{941--951}.
\newblock


\bibitem[\protect\citeauthoryear{Kenter, Borisov, Van~Gysel, Dehghani,
  de~Rijke, and Mitra}{Kenter et~al\mbox{.}}{2017}]%
        {Kenter2017nn4ir}
\bibfield{author}{\bibinfo{person}{Tom Kenter}, \bibinfo{person}{Alexey
  Borisov}, \bibinfo{person}{Christophe Van~Gysel}, \bibinfo{person}{Mostafa
  Dehghani}, \bibinfo{person}{Maarten de Rijke}, {and} \bibinfo{person}{Bhaskar
  Mitra}.} \bibinfo{year}{2017}\natexlab{}.
\newblock \showarticletitle{Neural networks for information retrieval}. In
  \bibinfo{booktitle}{\emph{SIGIR 2017}}. ACM, \bibinfo{pages}{1403--1406}.
\newblock


\bibitem[\protect\citeauthoryear{Kenter and de~Rijke}{Kenter and
  de~Rijke}{2015}]%
        {Kenter2015shorttext}
\bibfield{author}{\bibinfo{person}{Tom Kenter} {and} \bibinfo{person}{Maarten
  de Rijke}.} \bibinfo{year}{2015}\natexlab{}.
\newblock \showarticletitle{Short text similarity with word embeddings}. In
  \bibinfo{booktitle}{\emph{CIKM}}. ACM, \bibinfo{pages}{1411--1420}.
\newblock


\bibitem[\protect\citeauthoryear{Kingma and Ba}{Kingma and Ba}{2014}]%
        {Kingma2014}
\bibfield{author}{\bibinfo{person}{Diederik~P. Kingma} {and}
  \bibinfo{person}{Jimmy Ba}.} \bibinfo{year}{2014}\natexlab{}.
\newblock \showarticletitle{Adam: {A} Method for Stochastic Optimization}.
\newblock \bibinfo{journal}{\emph{CoRR}}  \bibinfo{volume}{abs/1412.6980}
  (\bibinfo{year}{2014}).
\newblock


\bibitem[\protect\citeauthoryear{Krizhevsky, Sutskever, and Hinton}{Krizhevsky
  et~al\mbox{.}}{2012}]%
        {Krizhevsky2012net}
\bibfield{author}{\bibinfo{person}{Alex Krizhevsky}, \bibinfo{person}{Ilya
  Sutskever}, {and} \bibinfo{person}{Geoffrey~E Hinton}.}
  \bibinfo{year}{2012}\natexlab{}.
\newblock \showarticletitle{Imagenet classification with deep convolutional
  neural networks}. In \bibinfo{booktitle}{\emph{NIPS}}.
  \bibinfo{pages}{1097--1105}.
\newblock


\bibitem[\protect\citeauthoryear{Le and Mikolov}{Le and Mikolov}{2014}]%
        {Le2014}
\bibfield{author}{\bibinfo{person}{Quoc Le} {and} \bibinfo{person}{Tomas
  Mikolov}.} \bibinfo{year}{2014}\natexlab{}.
\newblock \showarticletitle{Distributed representations of sentences and
  documents}. In \bibinfo{booktitle}{\emph{ICML}}. \bibinfo{pages}{1188--1196}.
\newblock


\bibitem[\protect\citeauthoryear{LeCun, Bottou, Bengio, and Haffner}{LeCun
  et~al\mbox{.}}{1998}]%
        {LeCun1998gradient}
\bibfield{author}{\bibinfo{person}{Yann LeCun}, \bibinfo{person}{L{\'e}on
  Bottou}, \bibinfo{person}{Yoshua Bengio}, {and} \bibinfo{person}{Patrick
  Haffner}.} \bibinfo{year}{1998}\natexlab{}.
\newblock \showarticletitle{Gradient-based learning applied to document
  recognition}.
\newblock \bibinfo{journal}{\emph{IEEE}} \bibinfo{volume}{86},
  \bibinfo{number}{11} (\bibinfo{year}{1998}), \bibinfo{pages}{2278--2324}.
\newblock


\bibitem[\protect\citeauthoryear{Levy and Goldberg}{Levy and Goldberg}{2014}]%
        {Levy2014neural}
\bibfield{author}{\bibinfo{person}{Omer Levy} {and} \bibinfo{person}{Yoav
  Goldberg}.} \bibinfo{year}{2014}\natexlab{}.
\newblock \showarticletitle{Neural word embedding as implicit matrix
  factorization}. In \bibinfo{booktitle}{\emph{NIPS}}.
  \bibinfo{pages}{2177--2185}.
\newblock


\bibitem[\protect\citeauthoryear{Levy, Goldberg, and Dagan}{Levy
  et~al\mbox{.}}{2015}]%
        {Levy2015improving}
\bibfield{author}{\bibinfo{person}{Omer Levy}, \bibinfo{person}{Yoav Goldberg},
  {and} \bibinfo{person}{Ido Dagan}.} \bibinfo{year}{2015}\natexlab{}.
\newblock \showarticletitle{Improving distributional similarity with lessons
  learned from word embeddings}.
\newblock \bibinfo{journal}{\emph{TACL}}  \bibinfo{volume}{3}
  (\bibinfo{year}{2015}), \bibinfo{pages}{211--225}.
\newblock


\bibitem[\protect\citeauthoryear{Li and Xu}{Li and Xu}{2014}]%
        {Li2014}
\bibfield{author}{\bibinfo{person}{Hang Li} {and} \bibinfo{person}{Jun Xu}.}
  \bibinfo{year}{2014}\natexlab{}.
\newblock \showarticletitle{Semantic matching in search}.
\newblock \bibinfo{journal}{\emph{Foundations and Trends in Information
  Retrieval}} \bibinfo{volume}{7}, \bibinfo{number}{5} (\bibinfo{date}{June}
  \bibinfo{year}{2014}), \bibinfo{pages}{343--469}.
\newblock


\bibitem[\protect\citeauthoryear{Liu}{Liu}{2011}]%
        {Liu2011}
\bibfield{author}{\bibinfo{person}{Tie-Yan Liu}.}
  \bibinfo{year}{2011}\natexlab{}.
\newblock \bibinfo{booktitle}{\emph{Learning to Rank for Information
  Retrieval}}.
\newblock \bibinfo{publisher}{Springer}.
\newblock


\bibitem[\protect\citeauthoryear{Luhn}{Luhn}{1958}]%
        {Luhn1958significance}
\bibfield{author}{\bibinfo{person}{Hans~Peter Luhn}.}
  \bibinfo{year}{1958}\natexlab{}.
\newblock \showarticletitle{The automatic creation of literature abstracts}.
\newblock \bibinfo{journal}{\emph{IBM Journal of R\&D}}  \bibinfo{volume}{2}
  (\bibinfo{year}{1958}), \bibinfo{pages}{159--165}.
\newblock


\bibitem[\protect\citeauthoryear{Mikolov, Chen, Corrado, and Dean}{Mikolov
  et~al\mbox{.}}{2013a}]%
        {Mikolov2013compositionality}
\bibfield{author}{\bibinfo{person}{Tomas Mikolov}, \bibinfo{person}{Kai Chen},
  \bibinfo{person}{Greg Corrado}, {and} \bibinfo{person}{Jeffrey Dean}.}
  \bibinfo{year}{2013}\natexlab{a}.
\newblock \showarticletitle{Distributed representations of words and phrases
  and their compositionality}. In \bibinfo{booktitle}{\emph{NIPS}}.
  \bibinfo{pages}{3111--3119}.
\newblock


\bibitem[\protect\citeauthoryear{Mikolov, Corrado, Chen, and Dean}{Mikolov
  et~al\mbox{.}}{2013b}]%
        {Mikolov2013word2vec}
\bibfield{author}{\bibinfo{person}{Tomas Mikolov}, \bibinfo{person}{Greg
  Corrado}, \bibinfo{person}{Kai Chen}, {and} \bibinfo{person}{Jeffrey Dean}.}
  \bibinfo{year}{2013}\natexlab{b}.
\newblock \bibinfo{title}{Efficient estimation of word representations in
  vector space}.
\newblock \bibinfo{howpublished}{arXiv 1301.3781}.   (\bibinfo{year}{2013}),
  \bibinfo{numpages}{12}~pages.
\newblock


\bibitem[\protect\citeauthoryear{Mitra, Diaz, and Craswell}{Mitra
  et~al\mbox{.}}{2017}]%
        {Mitra2017distributed}
\bibfield{author}{\bibinfo{person}{Bhaskar Mitra}, \bibinfo{person}{Fernando
  Diaz}, {and} \bibinfo{person}{Nick Craswell}.}
  \bibinfo{year}{2017}\natexlab{}.
\newblock \showarticletitle{Learning tomatch using local and distributed
  representations of text for web search}. In \bibinfo{booktitle}{\emph{WWW}}.
\newblock


\bibitem[\protect\citeauthoryear{Moore}{Moore}{1998}]%
        {Moore1998cramming}
\bibfield{author}{\bibinfo{person}{Gordon~E Moore}.}
  \bibinfo{year}{1998}\natexlab{}.
\newblock \showarticletitle{Cramming more components onto integrated circuits}.
\newblock \bibinfo{journal}{\emph{Proc. IEEE}} \bibinfo{volume}{86},
  \bibinfo{number}{1} (\bibinfo{year}{1998}), \bibinfo{pages}{82--85}.
\newblock


\bibitem[\protect\citeauthoryear{Muja and Lowe}{Muja and Lowe}{2014}]%
        {muja2014scalableknn}
\bibfield{author}{\bibinfo{person}{Marius Muja} {and} \bibinfo{person}{David~G
  Lowe}.} \bibinfo{year}{2014}\natexlab{}.
\newblock \showarticletitle{Scalable nearest neighbor algorithms for high
  dimensional data}.
\newblock \bibinfo{journal}{\emph{Pattern Analysis and Machine Intelligence}}
  \bibinfo{volume}{36}, \bibinfo{number}{11} (\bibinfo{year}{2014}),
  \bibinfo{pages}{2227--2240}.
\newblock


\bibitem[\protect\citeauthoryear{Nalisnick, Mitra, Craswell, and
  Caruana}{Nalisnick et~al\mbox{.}}{2016}]%
        {Nalisnick2016desm}
\bibfield{author}{\bibinfo{person}{Eric Nalisnick}, \bibinfo{person}{Bhaskar
  Mitra}, \bibinfo{person}{Nick Craswell}, {and} \bibinfo{person}{Rich
  Caruana}.} \bibinfo{year}{2016}\natexlab{}.
\newblock \showarticletitle{Improving document ranking with dual word
  embeddings}. In \bibinfo{booktitle}{\emph{WWW}}. International World Wide Web
  Conferences Steering Committee, \bibinfo{pages}{83--84}.
\newblock


\bibitem[\protect\citeauthoryear{Onal, Zhang, Altingovde, Rahman, Karagoz,
  Braylan, Dang, Chang, Kim, McNamara, Angert, Banner, Khetan, McDonnell,
  Nguyen, Xu, Wallace, de~Rijke, and Lease}{Onal et~al\mbox{.}}{2018}]%
        {Onal2017neural}
\bibfield{author}{\bibinfo{person}{Kezban~Dilek Onal}, \bibinfo{person}{Ye
  Zhang}, \bibinfo{person}{Ismail~Sengor Altingovde},
  \bibinfo{person}{Md~Mustafizur Rahman}, \bibinfo{person}{Pinar Karagoz},
  \bibinfo{person}{Alex Braylan}, \bibinfo{person}{Brandon Dang},
  \bibinfo{person}{Heng-Lu Chang}, \bibinfo{person}{Henna Kim},
  \bibinfo{person}{Quinten McNamara}, \bibinfo{person}{Aaron Angert},
  \bibinfo{person}{Edward Banner}, \bibinfo{person}{Vivek Khetan},
  \bibinfo{person}{Tyler McDonnell}, \bibinfo{person}{An~Thanh Nguyen},
  \bibinfo{person}{Dan Xu}, \bibinfo{person}{Byron~C. Wallace},
  \bibinfo{person}{Maarten de Rijke}, {and} \bibinfo{person}{Matthew Lease}.}
  \bibinfo{year}{2018}\natexlab{}.
\newblock \showarticletitle{Neural information retrieval: At the end of the
  early years}.
\newblock \bibinfo{journal}{\emph{Information Retrieval Journal}}
  (\bibinfo{year}{2018}).
\newblock
\newblock
\shownote{To appear.}


\bibitem[\protect\citeauthoryear{Pennington, Socher, and Manning}{Pennington
  et~al\mbox{.}}{2014}]%
        {Pennington2014}
\bibfield{author}{\bibinfo{person}{Jeffrey Pennington},
  \bibinfo{person}{Richard Socher}, {and} \bibinfo{person}{Christopher~D
  Manning}.} \bibinfo{year}{2014}\natexlab{}.
\newblock \showarticletitle{GloVe: Global vectors for word representation}. In
  \bibinfo{booktitle}{\emph{EMNLP}}. \bibinfo{pages}{1532--1543}.
\newblock


\bibitem[\protect\citeauthoryear{{\v R}eh{\r u}{\v r}ek and Sojka}{{\v R}eh{\r
  u}{\v r}ek and Sojka}{2010}]%
        {Rehurek2010gensim}
\bibfield{author}{\bibinfo{person}{Radim {\v R}eh{\r u}{\v r}ek} {and}
  \bibinfo{person}{Petr Sojka}.} \bibinfo{year}{2010}\natexlab{}.
\newblock \showarticletitle{Software framework for topic modelling with large
  corpora}. In \bibinfo{booktitle}{\emph{{Proceedings of the LREC 2010 Workshop
  on New Challenges for NLP Frameworks}}}. \bibinfo{publisher}{ELRA},
  \bibinfo{address}{Valletta, Malta}, \bibinfo{pages}{45--50}.
\newblock
\newblock
\shownote{\url{http://is.muni.cz/publication/884893/en}.}


\bibitem[\protect\citeauthoryear{Robertson}{Robertson}{2004}]%
        {Robertson2004idf}
\bibfield{author}{\bibinfo{person}{Stephen Robertson}.}
  \bibinfo{year}{2004}\natexlab{}.
\newblock \showarticletitle{Understanding inverse document frequency: on
  theoretical arguments for IDF}.
\newblock \bibinfo{journal}{\emph{Journal of documentation}}
  \bibinfo{volume}{60}, \bibinfo{number}{5} (\bibinfo{year}{2004}),
  \bibinfo{pages}{503--520}.
\newblock


\bibitem[\protect\citeauthoryear{Robertson and Walker}{Robertson and
  Walker}{1994}]%
        {Robertson1994bm25}
\bibfield{author}{\bibinfo{person}{Stephen~E Robertson} {and}
  \bibinfo{person}{Steve Walker}.} \bibinfo{year}{1994}\natexlab{}.
\newblock \showarticletitle{Some simple effective approximations to the
  2-poisson model for probabilistic weighted retrieval}. In
  \bibinfo{booktitle}{\emph{SIGIR}}. \bibinfo{pages}{232--241}.
\newblock


\bibitem[\protect\citeauthoryear{Sak, Senior, and Beaufays}{Sak
  et~al\mbox{.}}{2014}]%
        {Sak2015acousticlstm}
\bibfield{author}{\bibinfo{person}{Hasim Sak}, \bibinfo{person}{Andrew~W
  Senior}, {and} \bibinfo{person}{Fran{\c{c}}oise Beaufays}.}
  \bibinfo{year}{2014}\natexlab{}.
\newblock \showarticletitle{Long short-term memory recurrent neural network
  architectures for large scale acoustic modeling}. In
  \bibinfo{booktitle}{\emph{Interspeech}}. \bibinfo{pages}{338--342}.
\newblock


\bibitem[\protect\citeauthoryear{Salakhutdinov and Hinton}{Salakhutdinov and
  Hinton}{2009}]%
        {Salakhutdinov2009}
\bibfield{author}{\bibinfo{person}{Ruslan Salakhutdinov} {and}
  \bibinfo{person}{Geoffrey Hinton}.} \bibinfo{year}{2009}\natexlab{}.
\newblock \showarticletitle{Semantic hashing}.
\newblock \bibinfo{journal}{\emph{Int. J. Approximate Reasoning}}
  \bibinfo{volume}{50}, \bibinfo{number}{7} (\bibinfo{year}{2009}),
  \bibinfo{pages}{969--978}.
\newblock


\bibitem[\protect\citeauthoryear{Shaw, Fox, Shaw, and Fox}{Shaw
  et~al\mbox{.}}{1994}]%
        {Shaw1994}
\bibfield{author}{\bibinfo{person}{Joseph~A. Shaw}, \bibinfo{person}{Edward~A.
  Fox}, \bibinfo{person}{Joseph~A. Shaw}, {and} \bibinfo{person}{Edward~A.
  Fox}.} \bibinfo{year}{1994}\natexlab{}.
\newblock \showarticletitle{Combination of multiple searches}. In
  \bibinfo{booktitle}{\emph{TREC}}. \bibinfo{pages}{243--252}.
\newblock


\bibitem[\protect\citeauthoryear{Shen, He, Gao, Deng, and Mesnil}{Shen
  et~al\mbox{.}}{2014}]%
        {Shen2014}
\bibfield{author}{\bibinfo{person}{Yelong Shen}, \bibinfo{person}{Xiaodong He},
  \bibinfo{person}{Jianfeng Gao}, \bibinfo{person}{Li Deng}, {and}
  \bibinfo{person}{Gr\'{e}goire Mesnil}.} \bibinfo{year}{2014}\natexlab{}.
\newblock \showarticletitle{A latent semantic model with convolutional-pooling
  structure for information retrieval}. In \bibinfo{booktitle}{\emph{CIKM}}.
  \bibinfo{pages}{101--110}.
\newblock


\bibitem[\protect\citeauthoryear{Smucker, Allan, and Carterette}{Smucker
  et~al\mbox{.}}{2007}]%
        {Smucker2007}
\bibfield{author}{\bibinfo{person}{Mark~D Smucker}, \bibinfo{person}{James
  Allan}, {and} \bibinfo{person}{Ben Carterette}.}
  \bibinfo{year}{2007}\natexlab{}.
\newblock \showarticletitle{A comparison of statistical significance tests for
  information retrieval evaluation}. In \bibinfo{booktitle}{\emph{CIKM}}. ACM,
  \bibinfo{pages}{623--632}.
\newblock


\bibitem[\protect\citeauthoryear{Sparck~Jones}{Sparck~Jones}{1972}]%
        {Sparck1972specificity}
\bibfield{author}{\bibinfo{person}{Karen Sparck~Jones}.}
  \bibinfo{year}{1972}\natexlab{}.
\newblock \showarticletitle{A statistical interpretation of term specificity
  and its application in retrieval}.
\newblock \bibinfo{journal}{\emph{Journal of documentation}}
  \bibinfo{volume}{28}, \bibinfo{number}{1} (\bibinfo{year}{1972}),
  \bibinfo{pages}{11--21}.
\newblock


\bibitem[\protect\citeauthoryear{Strohman, Metzler, Turtle, and Croft}{Strohman
  et~al\mbox{.}}{2005}]%
        {Strohman2005indri}
\bibfield{author}{\bibinfo{person}{Trevor Strohman}, \bibinfo{person}{Donald
  Metzler}, \bibinfo{person}{Howard Turtle}, {and} \bibinfo{person}{W~Bruce
  Croft}.} \bibinfo{year}{2005}\natexlab{}.
\newblock \showarticletitle{Indri: A language model-based search engine for
  complex queries}. In \bibinfo{booktitle}{\emph{ICIA}}.
\newblock


\bibitem[\protect\citeauthoryear{Sutskever, Vinyals, and Le}{Sutskever
  et~al\mbox{.}}{2014}]%
        {Sutskever2014seq2seq}
\bibfield{author}{\bibinfo{person}{Ilya Sutskever}, \bibinfo{person}{Oriol
  Vinyals}, {and} \bibinfo{person}{Quoc~V Le}.}
  \bibinfo{year}{2014}\natexlab{}.
\newblock \showarticletitle{Sequence to sequence learning with neural
  networks}. In \bibinfo{booktitle}{\emph{NIPS}}. \bibinfo{pages}{3104--3112}.
\newblock


\bibitem[\protect\citeauthoryear{Tran, Bisazza, and Monz}{Tran
  et~al\mbox{.}}{2016}]%
        {Tran2016rmn}
\bibfield{author}{\bibinfo{person}{Ke Tran}, \bibinfo{person}{Arianna Bisazza},
  {and} \bibinfo{person}{Christof Monz}.} \bibinfo{year}{2016}\natexlab{}.
\newblock \showarticletitle{Recurrent memory network for language modeling}. In
  \bibinfo{booktitle}{\emph{NAACL-HLT}}. \bibinfo{publisher}{ACL},
  \bibinfo{pages}{321--331}.
\newblock


\bibitem[\protect\citeauthoryear{TREC}{TREC}{1999}]%
        {TRECAdhoc}
\bibfield{author}{\bibinfo{person}{TREC}.}
  \bibinfo{year}{1992--1999}\natexlab{}.
\newblock \bibinfo{title}{{TREC1-8 Adhoc Track}}.
\newblock \bibinfo{howpublished}{\url{http://trec.nist.gov/data/qrels_eng}}.
  (\bibinfo{year}{1992--1999}).
\newblock


\bibitem[\protect\citeauthoryear{Tu, Huang, Luo, and He}{Tu
  et~al\mbox{.}}{2016}]%
        {Tu2016semantic}
\bibfield{author}{\bibinfo{person}{Xinhui Tu}, \bibinfo{person}{Jimmy~Xiangji
  Huang}, \bibinfo{person}{Jing Luo}, {and} \bibinfo{person}{Tingting He}.}
  \bibinfo{year}{2016}\natexlab{}.
\newblock \showarticletitle{Exploiting Semantic Coherence Features for
  Information Retrieval}. In \bibinfo{booktitle}{\emph{SIGIR}}.
  \bibinfo{publisher}{ACM}, \bibinfo{pages}{837--840}.
\newblock


\bibitem[\protect\citeauthoryear{Turian, Ratinov, and Bengio}{Turian
  et~al\mbox{.}}{2010}]%
        {Turian2010representations}
\bibfield{author}{\bibinfo{person}{Joseph Turian}, \bibinfo{person}{Lev
  Ratinov}, {and} \bibinfo{person}{Yoshua Bengio}.}
  \bibinfo{year}{2010}\natexlab{}.
\newblock \showarticletitle{Word representations: a simple and general method
  for semi-supervised learning}. In \bibinfo{booktitle}{\emph{ACL}}.
  Association for Computational Linguistics, \bibinfo{pages}{384--394}.
\newblock


\bibitem[\protect\citeauthoryear{Van~Gysel, de~Rijke, and Kanoulas}{Van~Gysel
  et~al\mbox{.}}{2016a}]%
        {VanGysel2016products}
\bibfield{author}{\bibinfo{person}{Christophe Van~Gysel},
  \bibinfo{person}{Maarten de Rijke}, {and} \bibinfo{person}{Evangelos
  Kanoulas}.} \bibinfo{year}{2016}\natexlab{a}.
\newblock \showarticletitle{Learning latent vector spaces for product search}.
  In \bibinfo{booktitle}{\emph{CIKM}}. \bibinfo{publisher}{ACM},
  \bibinfo{pages}{165--174}.
\newblock


\bibitem[\protect\citeauthoryear{Van~Gysel, de~Rijke, and Kanoulas}{Van~Gysel
  et~al\mbox{.}}{2017a}]%
        {VanGysel2017pyndri}
\bibfield{author}{\bibinfo{person}{Christophe Van~Gysel},
  \bibinfo{person}{Maarten de Rijke}, {and} \bibinfo{person}{Evangelos
  Kanoulas}.} \bibinfo{year}{2017}\natexlab{a}.
\newblock \showarticletitle{Pyndri: A Python interface to the Indri search
  engine}. In \bibinfo{booktitle}{\emph{ECIR}}. \bibinfo{publisher}{Springer},
  \bibinfo{pages}{744--748}.
\newblock


\bibitem[\protect\citeauthoryear{Van~Gysel, de~Rijke, and Kanoulas}{Van~Gysel
  et~al\mbox{.}}{2017b}]%
        {VanGysel2017sert}
\bibfield{author}{\bibinfo{person}{Christophe Van~Gysel},
  \bibinfo{person}{Maarten de Rijke}, {and} \bibinfo{person}{Evangelos
  Kanoulas}.} \bibinfo{year}{2017}\natexlab{b}.
\newblock \showarticletitle{Semantic entity retrieval toolkit}. In
  \bibinfo{booktitle}{\emph{Neu-IR 2017}}.
\newblock


\bibitem[\protect\citeauthoryear{Van~Gysel, de~Rijke, and Worring}{Van~Gysel
  et~al\mbox{.}}{2016b}]%
        {VanGysel2016experts}
\bibfield{author}{\bibinfo{person}{Christophe Van~Gysel},
  \bibinfo{person}{Maarten de Rijke}, {and} \bibinfo{person}{Marcel Worring}.}
  \bibinfo{year}{2016}\natexlab{b}.
\newblock \showarticletitle{Unsupervised, efficient and semantic expertise
  retrieval}. In \bibinfo{booktitle}{\emph{WWW}}. \bibinfo{publisher}{ACM},
  \bibinfo{pages}{1069--1079}.
\newblock


\bibitem[\protect\citeauthoryear{Van~Gysel, Mitra, Venanzi, Rosemarin, Kukla,
  Grudzien, and Cancedda}{Van~Gysel et~al\mbox{.}}{2017c}]%
        {VanGysel2017replywith}
\bibfield{author}{\bibinfo{person}{Christophe Van~Gysel},
  \bibinfo{person}{Bhaskar Mitra}, \bibinfo{person}{Matteo Venanzi},
  \bibinfo{person}{Roy Rosemarin}, \bibinfo{person}{Grzegorz Kukla},
  \bibinfo{person}{Piotr Grudzien}, {and} \bibinfo{person}{Nicola Cancedda}.}
  \bibinfo{year}{2017}\natexlab{c}.
\newblock \showarticletitle{Reply with: Proactive recommendation of email
  attachments}. In \bibinfo{booktitle}{\emph{CIKM}}. \bibinfo{publisher}{ACM},
  \bibinfo{pages}{327--336}.
\newblock


\bibitem[\protect\citeauthoryear{van Rijsbergen}{van Rijsbergen}{1979}]%
        {Rijsbergen1979}
\bibfield{author}{\bibinfo{person}{C.~J. van Rijsbergen}.}
  \bibinfo{year}{1979}\natexlab{}.
\newblock \bibinfo{booktitle}{\emph{Information Retrieval}
  (\bibinfo{edition}{2nd} ed.)}.
\newblock \bibinfo{publisher}{Butterworth-Heinemann}.
\newblock


\bibitem[\protect\citeauthoryear{Vidal-Naquet and Ullman}{Vidal-Naquet and
  Ullman}{2003}]%
        {Vidal2003sift}
\bibfield{author}{\bibinfo{person}{Michel Vidal-Naquet} {and}
  \bibinfo{person}{Shimon Ullman}.} \bibinfo{year}{2003}\natexlab{}.
\newblock \showarticletitle{Object recognition with informative features and
  linear classification}. In \bibinfo{booktitle}{\emph{ICCV}}. IEEE,
  \bibinfo{pages}{281}.
\newblock


\bibitem[\protect\citeauthoryear{Voorhees}{Voorhees}{2005}]%
        {Voorhees2005robust}
\bibfield{author}{\bibinfo{person}{Ellen~M. Voorhees}.}
  \bibinfo{year}{2005}\natexlab{}.
\newblock \showarticletitle{The {TREC} Robust Retrieval Track}.
\newblock \bibinfo{journal}{\emph{SIGIR Forum}} \bibinfo{volume}{39},
  \bibinfo{number}{1} (\bibinfo{date}{June} \bibinfo{year}{2005}),
  \bibinfo{pages}{11--20}.
\newblock


\bibitem[\protect\citeauthoryear{Vuli{\'c} and Moens}{Vuli{\'c} and
  Moens}{2015}]%
        {Vulic2015monolingual}
\bibfield{author}{\bibinfo{person}{Ivan Vuli{\'c}} {and}
  \bibinfo{person}{Marie-Francine Moens}.} \bibinfo{year}{2015}\natexlab{}.
\newblock \showarticletitle{Monolingual and cross-lingual information retrieval
  models based on (bilingual) word embeddings}. In
  \bibinfo{booktitle}{\emph{SIGIR}}. ACM, \bibinfo{pages}{363--372}.
\newblock


\bibitem[\protect\citeauthoryear{Wei and Croft}{Wei and Croft}{2006}]%
        {Wei2006lda}
\bibfield{author}{\bibinfo{person}{Xing Wei} {and} \bibinfo{person}{W~Bruce
  Croft}.} \bibinfo{year}{2006}\natexlab{}.
\newblock \showarticletitle{LDA-based document models for ad-hoc retrieval}. In
  \bibinfo{booktitle}{\emph{SIGIR}}. ACM, \bibinfo{pages}{178--185}.
\newblock


\bibitem[\protect\citeauthoryear{Wikipedia}{Wikipedia}{2017}]%
        {Wiki2017nvidia}
\bibfield{author}{\bibinfo{person}{Wikipedia}.}
  \bibinfo{year}{2017}\natexlab{}.
\newblock \bibinfo{title}{List of Nvidia graphics processing units ---
  Wikipedia{,} The Free Encyclopedia}.
\newblock   (\bibinfo{year}{2017}).
\newblock
\urldef\tempurl%
\url{https://en.wikipedia.org/w/index.php?title=List_of_Nvidia_graphics_processing_units&oldid=792964538}
\showURL{%
\tempurl}
\newblock
\shownote{[Online; accessed 8-August-2017].}


\bibitem[\protect\citeauthoryear{Zamani and Croft}{Zamani and Croft}{2016a}]%
        {Zamani2016embeddinglm}
\bibfield{author}{\bibinfo{person}{Hamed Zamani} {and}
  \bibinfo{person}{W.~Bruce Croft}.} \bibinfo{year}{2016}\natexlab{a}.
\newblock \showarticletitle{Embedding-based query language models}. In
  \bibinfo{booktitle}{\emph{ICTIR}}. \bibinfo{publisher}{ACM},
  \bibinfo{pages}{147--156}.
\newblock


\bibitem[\protect\citeauthoryear{Zamani and Croft}{Zamani and Croft}{2016b}]%
        {Zamani2016queryexpansion}
\bibfield{author}{\bibinfo{person}{Hamed Zamani} {and}
  \bibinfo{person}{W.~Bruce Croft}.} \bibinfo{year}{2016}\natexlab{b}.
\newblock \showarticletitle{Estimating embedding vectors for queries}. In
  \bibinfo{booktitle}{\emph{ICTIR}}. \bibinfo{publisher}{ACM},
  \bibinfo{pages}{123--132}.
\newblock


\bibitem[\protect\citeauthoryear{Zhai and Lafferty}{Zhai and Lafferty}{2004}]%
        {Zhai2004smoothing}
\bibfield{author}{\bibinfo{person}{Chengxiang Zhai} {and} \bibinfo{person}{John
  Lafferty}.} \bibinfo{year}{2004}\natexlab{}.
\newblock \showarticletitle{A study of smoothing methods for language models
  applied to information retrieval}.
\newblock \bibinfo{journal}{\emph{TOIS}} \bibinfo{volume}{22},
  \bibinfo{number}{2} (\bibinfo{year}{2004}), \bibinfo{pages}{179--214}.
\newblock


\bibitem[\protect\citeauthoryear{Zuccon, Koopman, Bruza, and Azzopardi}{Zuccon
  et~al\mbox{.}}{2015}]%
        {Zuccon2015nntm}
\bibfield{author}{\bibinfo{person}{Guido Zuccon}, \bibinfo{person}{Bevan
  Koopman}, \bibinfo{person}{Peter Bruza}, {and} \bibinfo{person}{Leif
  Azzopardi}.} \bibinfo{year}{2015}\natexlab{}.
\newblock \showarticletitle{Integrating and evaluating neural word embeddings
  in information retrieval}. In \bibinfo{booktitle}{\emph{ADCS}}.
  \bibinfo{publisher}{ACM}, Article \bibinfo{articleno}{12},
  \bibinfo{numpages}{8}~pages.
\newblock


\end{thebibliography}

\end{document}